\documentclass[a4paper,12pt]{article}
\pdfoutput=1 

\usepackage{jheppub} 

\usepackage[T1]{fontenc} 
\usepackage{amsmath,amssymb}
\usepackage{slashed}
\newcommand{\be}{\begin{equation}}
\newcommand{\ee}{\end{equation}}
\newcommand{\ben}{\begin{eqnarray}}
\newcommand{\een}{\end{eqnarray}}
\newcommand{\nn}{\nonumber}
\newcommand{\p}{\partial}
\def\l{\left}
\def\r{\right}
\DeclareMathOperator{\sech}{sech}

\title{\bf{Localization on $AdS_2\times S^1$ }}
\author[a]{\small{Justin R.  David},}
\author[b]{\small{Edi Gava},}
\author[c]{\small{Rajesh Kumar Gupta},}
\author[c]{\small{Kumar Narain}}
\affiliation[a]{\small{Centre for High Energy Physics, Indian Institute of Science,\\
C. V. Raman Avenue, Bangalore 560012, India.}}
\affiliation[b]{\small{INFN, sezione di Trieste, Italy}}
\affiliation[c]{\small{ICTP, Strada Costiera 11, 34151 Trieste, Italy}}

\emailAdd{ justin@cts.iisc.ernet.in}
\emailAdd{gava@ictp.it}
\emailAdd{rgupta@ictp.it}
\emailAdd{narain@ictp.it}

\abstract{
Conformal symmetry relates the metric on 
$AdS_2 \times S^{1}$  to that of  $S^3$.
This implies that under a   suitable choice of boundary conditions  for fields 
on $AdS_2$  the partition function of conformal field theories on 
these spaces must agree which makes 
 $AdS_2 \times S^{1}$  a  good testing ground  to study 
localization on non-compact spaces. 
We study supersymmetry on $AdS_2\times S^1$    and 
determine the localizing Lagrangian  for  ${\cal N}=2$ supersymmetric 
Chern-Simons theory on $AdS_2\times S^1$. 
We evaluate the partition function of ${\cal N}=2$ supersymmetric Chern-Simons theory 
on $AdS_2 \times S^1$ using localization, where the radius of  $S^1$ is $q$ times that of 
 $AdS_2$. 
With boundary conditions on $AdS_2\times S^1$ which 
ensure that all the physical  fields are normalizable and lie in the space 
of square integrable wave functions in $AdS_2$, 
 the result  for the partition function precisely agrees with that of the theory 
on the $q$-fold covering of $S^3$.   
} 

\begin{document} 
\maketitle
\flushbottom

\section{Introduction}

The method of localization 
  is a powerful technique to evaluate   observables 
in supersymmetric quantum field theories exactly. 
The method was first introduced in \cite{Witten:1992xu} and 
developed in \cite{Nekrasov:2003af,Nekrasov:2003rj}. 
It was revived by the work of \cite{Pestun:2007rz} which has led to 
the exact computations in supersymmetric 
quantum field theories in various dimensions and  manifolds.
Many of these exact computations have been used to provide
highly non-trivial  checks of the AdS/CFT correspondence. See 
the review \cite{Pestun:2016zxk} for a comprehensive list of references. 

Most of the activity has been focussed on  supersymmetric quantum field theory 
defined on curved but compact spaces. 
The main reason is because the localizing Lagrangian one adds 
is  exact  under the Fermionic symmetry $Q$  up to boundary terms. 
Therefore all such terms can be neglected on compact spaces without a 
boundary. 
Rigid supersymmetric  quantum field theories can also be
defined on curved  but non-compact spaces 
\cite{Festuccia:2011ws,Dumitrescu:2012ha,Klare:2013dka,Alday:2013lba,Closset:2013vra,Aharony:2015hix}. 
Quantum field theories on spaces of the form $AdS_n\times S^m$ 
are relevant  in  evaluating black hole entropy as well 
as  entanglement entropy across spherical entangling surfaces in 
conformal field theories. 
In this context, localization of ${\cal N}=2$ supergravity was studied 
in a  series of work on $AdS_2\times S^2$ to obtain black hole entropy 
of extremal black holes \cite{Dabholkar:2010uh,Dabholkar:2011ec,Gupta:2012cy,Dabholkar:2014ema,Gupta:2015gga,Murthy:2015yfa}. 
Localization of  supergravity on $AdS_4$ was also studied in the context of 
evaluating the quantum partition function in the bulk for the ABJM theory 
\cite{Dabholkar:2014wpa}.

Let us examine the second instance where   partition functions of quantum field theories 
on Anti-de Sitter spaces are important. 
The R\'{e}nyi entropy   of order $q$ of a spherical entangling surface 
in a  $d$ dimensional conformal field theory can be mapped 
to  the evaluation of the  partition function of the theory on a $q$-fold 
covering of the sphere $S^d$. This partition function is in turn related to  
the thermal partition function of the  conformal field theory on 
$AdS_{d-1} \times S^1$ where the radius of $S^1$ is $q$ times that of 
$AdS_{d}$ \cite{Casini:2011kv}.   
This relation between the partition functions on these 
surfaces offers a situation in which any  formulation  of 
localization  on  non-compact manifolds  is more controlled and  can be precisely checked. 
In the context of localization such  issues were 
previously explored in \cite{Huang:2014gca,Huang:2014pda}.

To be more concrete  we  focus our attention to conformal field theories 
in $d=3$. 
Consider the  following metric on the $3$-sphere
\begin{equation} \label{branched} 
ds^2_A = L^2 \left( \cos^2\phi d \tilde \tau^2 + d\phi^2 + \sin^2 \phi d \theta^2 \right) , 
\end{equation}
where the coordinates  take values  from 
\begin{equation}
 0\leq \tilde\tau \leq 2\pi q , \qquad 0 \leq \theta \leq 2\pi, \qquad 0 \leq \phi\leq \frac{\pi}{2}.
\end{equation}
When $q$ takes values in the set of positive integers, the metric in (\ref{branched})  
is that of  a $q$-fold covering of $S^3$ branched on the circle at $\phi = \pi/2$. 
Lets denote this space  with the metric in (\ref{branched}) by $A_q$.
Under the transformation 
\begin{equation}
 \sinh r = \tan \phi, 
\end{equation}
the metric in (\ref{branched}) is conformal related to 
\begin{equation}
 ds^2_B = L^2 ( d\tilde \tau^2 + dr^2 + \sinh^2 r d\theta^2),  
\end{equation}
where the coordinates take values from 
\begin{equation}
 0\leq \tilde \tau \leq 2\pi q , \qquad 0 \leq \theta \leq 2\pi, \qquad 0 \leq r  <\infty . 
\end{equation}
We denote this space by $B_q$.
The relation  between the metrics is given by 
\begin{equation}
 ds^2_A = \cos^2 \phi ds^2_B. 
\end{equation}
Partition functions of conformal field theories  defined on  $A_q$ 
should equal to the partition function of the same theory defined on $B_q$ 
with suitably chosen boundary conditions. 
In \cite{Klebanov:2011uf} it was shown that 
with fields satisfying normalizable boundary conditions in $AdS_2$ and  a suitable
regularization, the  thermal partition function of 
of free conformal scalars and free  massless fermions on  $B_q$ agreed precisely with 
that on $A_q$. 

In this paper we would like to test this relationship between the 
partition functions on spaces $A_q$ and $B_q$ for interacting  theories. 
We will restrict our attention to supersymmetric partition functions obtained by  the 
technique of localization. 
The simplest example of a non-trivial  super conformal field theory in $3$ dimensions is that 
of pure ${\cal N}=2$ supersymmetric Chern-Simons gauge theory. 
This theory has the added simplification of the fact that the 
fermions and the auxillary scalar do not have kinetic terms and therefore
the supersymmetric partition function on $S^3$ is related to the bosonic Chern-Simons 
on $S^3$ upto a normalization factor. 
The supersymmetric partition function of Chern-Simons theory coupled 
to matter on $A_q$ was evaluated in \cite{Nishioka:2013haa}. 

We first  set up the supersymmetric transformations of the vector multiplet on 
$AdS_2\times S^1$, we then determine the localization Lagragian and the 
partition function on $B_q$ by evaluating the one-loop determinants using 
the index approach.  
  Then the indices are evaluated by 
explicitly solving the differential equations and counting the solutions 
which contribute to the  index. This method was first introduced 
in \cite{Cabo-Bizet:2014nia}. 
We discuss in detail the boundary conditions on $B_q$   which ensure that the 
space of functions for which the indices are evaluated lie in the 
space of normalizable functions in $AdS_2$. 
We show that with these boundary conditions the 
 supersymmetric partition function on $B_q$  is identical to the partition function 
 on $A_q$.  This in turn ensures that the  partition function of 
 super Chern-Simons theory 
 obtained by localization on $B_q$ agrees with the partition function of pure bosonic 
 Chern-Simons theory. Given this agreement, we further consider a family of non-singular 3-manifolds, labelled by a continuous parameter $s\in [0,1]$, which are conformally equivalent to AdS$_2\times S^1$ and show that the index and the partition function do not change. These 3-manifolds are defined using the conformal transformation that does not change the asymptotic boundary conditions on the fields.

 The organisation of this paper is as follows. 
 In the next section as a warm up we show that the partition function of the 
 Abelian Chern Simons theory on $AdS_2 \times S^1$ is independent of 
 $q$ and agrees precisely with that on the space $A_q$ which was obtained
 in \cite{Klebanov:2011uf}.  The analysis of this section will show what are
 the boundary conditions imposed on the fields in $AdS_2$  which 
 results in the agreement. 
 In section 3  we study  ${\cal N} = 2$ supersymmetry   for the gauge multiplet 
 on $AdS_2 \times S^1$. 
 We solve for the Killing spinors on $AdS_2 \times S^1$ and 
 obtain the supersymmetric transformation under which 
 ${\cal N}=2$   Chern-Simons Lagrangian is supersymmetric 
 and then obtain the localizing term. 
 In the section 4 we evaluate the supersymmetric partition function 
 on $B_q$ 
  by performing the one loop determinants 
 using the index method.  We  discuss the boundary conditions
 of the functions over which the index is evaluated in detail. 
 We show that the result of the partition function 
 coincides with the supersymmetric partition function on 
 $A_q$. In section 5 we evaluate the expectation value of a supersymmetric Wilson loop operator. In section 6 we consider a family of 3-manifolds which are conformal to AdS$_2\times$S$^1$ and show that the index does not change. 
 Finally in section 7 we conclude with the discussion of the implications
 of these results.

\section{Abelian Chern Simons theory on $AdS_2\times S^1$}\label{OneloopAbelian}

As a warm up we begin testing the relationship between the partition functions 
on theories defined on the space $A_q$  and $B_q$ by first considering 
the case of the abelian Chern-Simons theory on the space $B_q$. 
The action of this theory is given by 
\begin{equation}
S_{CS} = \frac{i\kappa}{4\pi} \int  d^3x \varepsilon^{\mu\nu\rho}A_\mu\p_\nu A_\rho. 
\end{equation}
Here $\varepsilon^{\mu\nu\rho}$ is a tensor density with $\varepsilon^{\tau r \theta}=1$
and is related to Levi Civita tensor $\epsilon^{\mu\nu\rho}$ by 
\begin{equation}
\epsilon^{\mu\nu\rho}=\frac{1}{\sqrt{g}}\varepsilon^{\mu\nu\rho}, 
\end{equation}
where $g$ is the determinant of the metric.
The partition function ${\cal Z}_q$  of this theory on $A_q$ was evaluated  
in Appendix C.   of \cite{Klebanov:2011uf} and it was shown that the result is independent
of $q$ and is given by 
\begin{equation}
\log { Z}_q = -\frac{1}{2} \log \kappa.
\end{equation}
It is indeed expected that  the Chern-Simons partition 
function is a topological invariant and therefore should be independent 
of $q$ .  However this result is more significant since 
the space $A_q$ is not smooth. 

Our goal is now to reproduce this dependence on  $\kappa$ by evaluating the 
partition function of the theory on $B_q$. 
For convenience we  rescale the co-ordinate
\begin{equation}
L \tilde \tau = \tau.
\end{equation}
Without losing any generality we also choose
\begin{equation}
q = \frac{1}{L}.
\end{equation}
Then  we obtain the metric
\begin{equation}\label{Lmetric}
ds^2 = d\tau^2 + L^2(  dr^2 + \sinh^2 r d\theta^2 ).
\end{equation}
Now the range of $\tau$ is given by 
\begin{equation}
0 \leq \tau \leq 2\pi 
\end{equation}
We use the covariant gauge 
\begin{equation}
\nabla^\mu A_\mu = 0 .
\end{equation}
The action including the ghosts then become
\begin{equation}
S_{\rm ghost} = \int d^3 x \sqrt{g}  \left ( - \bar c\,  \Box c + i b \nabla^\mu A_\mu  \right) . 
\end{equation}
Here $c$ is the fermionic ghost, while $b$ is the bosonic ghost. 
The  $\Box$ refers to the Laplacian of a massless scalar with the 
metric in (\ref{Lmetric}). 
The total action including the ghosts  is given by 
\begin{equation}
S_{\rm total} = \frac{i\kappa}{4\pi} \int  d^3x \varepsilon^{\mu\nu\rho}A_\mu\p_\nu A_\rho
+ \int d^3 x \sqrt{g}  \left ( - \bar c\,  \Box c + i b \nabla^\mu A_\mu \right) . 
\end{equation}
%
%
%
Substituting the total action into the path integral 
and performing the integral over both the bosonic and the fermionic ghosts 
we are left with the following partition function. 
\begin{equation}\label{pi}
 { Z} = ( {\rm det} \, \Box )  \int  D[A_\mu] \delta \left[  \nabla^\mu A_\mu \right] \exp \left( \frac{ i \kappa}{4} 
 \int d^3 x \varepsilon^{\mu\nu\rho} ( A_\mu \partial_\nu A_\rho )  \right) . 
\end{equation}
Note that now we need to perform the path integral over configurations of gauge fields such that
the covariant gauge condition is satisfied. 
We will find a suitable set of variables where this can be carried out. 
Before we proceed, 
it is useful to write down the explicit expression for the determinant of the Laplacian 
on $AdS_2 \times S^1$. 
The eigen functions of the Laplacian are given by 
\begin{equation}
 \Box \Phi_{(\lambda, l; n ) } = - \left(  \tilde \lambda^2   + n^2\right)  
 \Phi_{(\lambda, l; n ) }, \qquad\qquad   \tilde \lambda^2   = \frac{1}{L^2} ( \lambda^2 + \frac{1}{4} ) . 
\end{equation}
Here $\lambda, l $ labels the quantum numbers onf $AdS_2$ and $n$ labels the Kaluza-Klein modes on 
$S^1$. 
The wave function $\Phi_{(\lambda, l; n ) }$ are constructed using the the eigen functions of the 
scalar Laplacian on $AdS_2$ together with the Fourier mode on $S^1$. They are defined as
\begin{equation}
 \Phi_{(\lambda, l; n ) } (r, \theta, \tau) =\frac{1}{L}  g_{\lambda, l } (r, \theta) e^{ i n \tau}, 
\end{equation}
and $g_{\lambda, l }(r, \theta) $ are normalizable eigen functions on $AdS_2$ which are given for example in 
\cite{Banerjee:2010qc,Camporesi:1994ga}\footnote{See equation (2.10) of \cite{Banerjee:2010qc}.}. The eigen value  
$\lambda$ takes values from $0$ to $\infty$ while $\{l, n \} \in \mathbb{Z}$. 
Using the orthonormal properties of $f_{\lambda, l }$ we have 
\begin{equation}
 \int d^3 x \sqrt{g}  (\Phi_{(\lambda, l; n ) })^*  \Phi_{(\lambda', l'; n' ) } = 2\pi \delta( \lambda - \lambda') 
 \delta_{l, -l'} \delta_{n, -n'}. 
\end{equation}
These eigen functions satisfy the following properties near the origin and the at boundary of 
$AdS_2$. 
\begin{eqnarray}
 \lim_{r\rightarrow 0 } \Phi_{(\lambda, l ; n)}{ (r\, \theta, \tau )} &\sim&  r^{|l|} e^{ i ( l \theta + n \tau)}, 
 \\ \nonumber
 \lim_{r \rightarrow \infty} \Phi_{(\lambda, l ; n)}{ (r\, \theta, \tau )} & \sim&  e^{ - \frac{r}{2} \pm i \lambda r } 
 e^{ i ( l \theta + n \tau)}. 
\end{eqnarray}
Therefore the  $c$ ghosts are expanded in terms of normalizable functions in $AdS_2$. 
Using these eigen functions we can write down the following expression for the  determinant 
of the Laplacian. 
\begin{eqnarray} \label{jacob}
 \log ( \text{det}\,\Box )   
=  \sum_{n= -\infty}^{n=\infty} \int_0^\infty d\lambda \mu(\lambda) \log 
\left( \tilde\lambda^2  + n^2  \right) . 
\end{eqnarray}
Here $\mu(\lambda)$ is the density of states  which is given by 
\begin{equation}\label{density}
 \mu(\lambda) = \frac{1}{2\pi L^2}  \lambda \tanh (\pi \lambda) . 
\end{equation}
This expression has to be regularized.  We can adopt the regularization procedure 
given in \cite{Klebanov:2011uf}, but we will not need it explicitly.


To impose the delta function in the path integral of (\ref{pi}) it is useful to expand the
gauge field  $A_\mu$ in terms of a  complete basis. 
On $AdS_2$, the vector can be expanded as a gradient of a scalar as well as a transvese component. 
Furthermore there are discrete modes for the vector on $AdS_2$ \cite{Camporesi:1994ga}. 
\begin{equation}\label{vecads2exp}
A_m= c_i\p_m\Phi_i(r, \theta,\tau)+d_i \epsilon_{mn}\p^n\Phi_i(r, \theta,\tau)+f_j\p_m\Psi_j(r, \theta,\tau).
%
\end{equation}
Here $m \in \{r, \theta\}$ the coordinates of $AdS_2$. 
The repeated index over $i$ refers to integration over  the $AdS_2$ eigen value $\lambda$ and 
the sum over the angular momentum mode $l$ in $AdS_2$ together with the sum 
over the Kaluza-Klein mode $n$. 
For example
\begin{equation}
 c_i \Phi_i(r, \theta,\tau) = \sum_{l, n} \int_0^\infty  d\lambda  c_{\{\lambda, l; n \}  } \Phi_{(\lambda , l ;n )}. 
\end{equation}
Note that though there is a sum over the infinite number of angular momentum  modes $l$  for 
each value of $\lambda $, it will turn out that the integrand  (\ref{pi})  is independent of $l$ and therefore 
we can just sum over the density of states using the measure in (\ref{density}). 
for each value of $\lambda$ in evaluating the logarithm of the 
partition function.  From now on we keep track of only the integral over $\lambda$ and the 
sum over the Kaluza-Klein modes. 
The discrete modes in (\ref{vecads2exp}) will play and important role in our analysis. 
Let us  recall the discrete modes of a vector on $AdS_2$. 
There are defined in terms of non-normalizable zero modes of the scalar Laplacian on 
$AdS_2$ which are given by 
\begin{equation}
 \Psi_{\{l,n\}}(r, \theta ,\tau)=\frac{1}{\sqrt{ 2\pi |l|}}
\left(\frac{\sinh r}{1+\cosh r}\right)^{|l|}e^{il\theta}e^{in\tau},\quad l=\pm 1,\pm 2,......
\end{equation}
These modes satisfy 
\begin{equation}
 \Box_{AdS_2} \Psi_{\{l,n\}}( r, \theta, \tau)  =0. 
\end{equation}
Note that though these  functions are non-normalizable,  the gradient  of these scalars satisfy the 
condition 
\begin{equation}
 \int  dr d\theta \sqrt{g_{AdS_2} }\nabla_m  \Psi_{\{l,n\}} \nabla^m (\Psi_{\{l' ,n\}} )^*  = \delta_{l, l'} 
\end{equation}
Thus the summation over $j$  which occurs with the discrete modes in (\ref{vecads2exp}) 
refers to the double sum over the angular momentum 
modes $l$ and the Kaluza-Klein modes $n$. 
Since the gauge field $A_m$ is real we have the property 
\begin{equation}\label{conjugate}
 c_{\{\lambda; n \} }^* = c_{\{ \lambda; - n \} } , \qquad\qquad   d_{\{ \lambda; n \} }^* = d_{\{ \lambda; -n \} }, 
 \qquad f_{ \{ l ; n\}}^* = f_{\{-l; -n \}}. 
\end{equation}
Finally   since the $A_\tau$ is a scalar on $AdS_2$ we can expand it as 
\begin{equation}  \label{atexp}
A_\tau=e_i\Phi_i(r, \theta ,\tau). 
\end{equation}
From (\ref{vecads2exp}) and (\ref{atexp}) we see that the gauge fields
satisfy the following boundary conditions in $AdS_2$. 
\begin{eqnarray}
 & & \lim_{r\rightarrow 0}  A_{r}^{( \lambda, l ; n) }  \sim  r^{|l| - 1} e^{ i ( l \theta +  n \tau) }, 
 \qquad
  \lim_{r\rightarrow 0}   A_{\theta}^{ ( \lambda, l ; n) } \sim r^{|l|} e^{ i ( l \theta +  n \tau) },  \\ \nonumber
   && 
  \lim_{r\rightarrow 0} A_{\tau}^{ ( \lambda, l ; n) } \sim  r^{|l|} e^{ i ( l \theta +  n \tau) },  
  \\ \nonumber
  && \lim_{r\rightarrow \infty}   A_{r}^{ ( \lambda, l ; n) } 
   \sim e^{ - \frac{r}{2} \pm i \lambda r } e^{ i ( l \theta +  n \tau) },
   \qquad 
  \lim_{r\rightarrow \infty}  A_{\theta}^{ ( \lambda, l ; n) }
  \sim e^{\frac{r}{2} \pm i \lambda r } e^{ i ( l \theta +  n \tau) },  \\ \nonumber
  &&  \lim_{r\rightarrow \infty}   A_{\tau }^{ ( \lambda, l ; n) } 
  \sim e^{ - \frac{r}{2} \pm i \lambda r } e^{ i ( l \theta +  n \tau) }, 
\end{eqnarray}
Note that for the gauge fields in the $AdS_2$ direction
 the angular momentum runs from $l \in \{ \pm 1, \pm, 2, \cdots \}$.

Let us now  find the Jacobian involved in changing the integration over $A_\mu$ in (\ref{pi})  to the
Fourier coefficients $\{ c_{\{\lambda; n \} }, d_{\{ \lambda; n \} },  e _{\{ \lambda; n \} },  f_{ \{ l ; n\}} \}$. 
We start with the measure over $A_\mu$ which is defined with the normalization 
\be \label{defmes}
 \int [DA_\mu] \exp\left( -\int d^3 x  \sqrt{g}A_\mu A^\mu\right) =1 . 
 \ee
 On substituting the expansion given in (\ref{vecads2exp})  and  (\ref{atexp}) 
 into the exponent we obtain 
 \begin{eqnarray}
  & &\int d^3 x  \sqrt{g}A_\mu A^\mu = 
    4\pi \sum_{n=1}^\infty  \left[ \tilde\lambda^2   (  c_{\{\lambda; n \} } c_{\{\lambda; -  n \} } + 
    d_{\{\lambda; n \} } d_{\{\lambda; -  n \} } )  + e_{\{\lambda; n \} } e_{\{\lambda; -  n \} }  \right]
     \nonumber  \\  \nonumber 
   & &  \qquad \qquad\qquad  +  2 \pi \left[ \tilde \lambda^2    ( c_{\{\lambda;  0 \} } c_{\{\lambda;  0 \} } + 
    d_{\{\lambda; 0  \} } d_{\{\lambda;  0 \} })  + e_{\{\lambda; 0 \} } e_{\{\lambda;  0  \} }  \right]  
    +   \sum_{l, n} f_{l, n} f_{-l -n} .  \\ 
 \end{eqnarray}
 Here we have written down the expression for a given value of $\lambda$. 
 Now using this expansion in (\ref{defmes})  and changing variables we obtain  
 \begin{eqnarray}
 & &  \int  \prod_{n}[ d c_{\{\lambda; n \} } d ( d_{\{\lambda; n \} })      d  e_{\{\lambda; n \} }]
 \prod_{n, l } [d  f_{l, n} ] 
  {\cal J} \times  \\ \nonumber
  & & \exp \left\{   
  - 4\pi \sum_{n=1}^\infty  \left[ \tilde\lambda^2   (  c_{\{\lambda; n \} } c_{\{\lambda; -  n \} } + 
    d_{\{\lambda; n \} } d_{\{\lambda; -  n \} } )  + e_{\{\lambda; n \} } e_{\{\lambda; -  n \} }  \right]
     \right.     \\  \nonumber
   & &  \left. -  2 \pi \left[ \tilde \lambda^2    ( c_{\{\lambda;  0 \} } c_{\{\lambda;  0 \} } + 
    d_{\{\lambda; 0  \} } d_{\{\lambda;  0 \} })  + e_{\{\lambda; 0 \} } e_{\{\lambda;  0  \} }  \right]  
    -   \sum_{l, n} f_{l, n} f_{-l -n}  \right\}  \\ \nonumber
  & & = 1, 
 \end{eqnarray}
where ${\cal J}$ is the Jacobian involved in the change of integration variables. 
Again we have written this only for a given value of $\lambda$. 
Peforming the integrations and taking into account of the density of states 
(\ref{density}) we obtain
\begin{equation} \label{jacob1}
 \log {\cal J}    =   \int d\lambda \mu(\lambda)  \left( 
 \log ( \tilde \lambda^2 )  + 2 \sum_{n=1}^\infty \log (\tilde \lambda^2)     \right) . 
\end{equation}
We have also used the relation (\ref{conjugate}) and the fact that  for $n=0$, the Fourier coefficients
are real.  Of course one needs to regularize the above expression, in fact 
the sum over $n$ can be done by using the $\zeta$ function regularization. 
At present we will assume a definite regularization has been chosen and proceed. 
We now rewrite  the delta function in terms of the Fourier coefficients. 
The divergence on $A_\mu$ can be written as
\begin{equation}
 \nabla^\mu A_\mu = c_i\Box\Phi_i(r, \theta,\tau)+e_i\p_\tau\Phi_i(r, \theta,\tau). 
\end{equation}
Therefore the delta function which imposes the transversality condition can be written as 
\begin{equation} \label{transdelt}
 \delta ( \nabla^\mu A_\mu) =  \prod_{ \lambda; n\neq 0 } \left[  \delta(  i n e_{\{ \lambda; n \} } - 
 \tilde\lambda^2  c_{\{ \lambda,  n \} }  \right]  
 \prod_{\lambda} [
 \delta( - \tilde \lambda^2 c_{ \{ \lambda , 0\} }  ) ]. 
\end{equation}
To arrive at this for the $n=0$ case we have used the fact that 
there are no normalizable scalars on $AdS_2$ with zero eigen value for the Laplacian. 
Now it is easy to see that performing the integration over $c_{\lambda, n}$ results in  
the Jacobian  \footnote{One can also perform the integration over  the variables $e_{\lambda, n}$  and arrive 
at the same final result. }
\begin{equation} \label{jacob2}
 \log {\hat{\cal  J } } = - \int d\lambda \mu(\lambda)  \left( 
 \log ( \tilde \lambda^2 )  + 2 \sum_{n=1}^\infty \log (\tilde \lambda^2)     \right) . 
\end{equation}
Note the factor of $2$ for the modes $n\neq 0$ results from the positive and negative 
Kaluz-Klein modes.  Comparing (\ref{jacob1}) and (\ref{jacob2}) 
 we see that the  Jacobian resulting from the change of variables precisely  cancels
on peforming the integral over  $c_{\lambda, n}$  using the delta function 
( \ref{transdelt} )which 
imposes the transversality condition.

 To summarize the 
 set of integration variables  left over are   $\left\{e_{\{\lambda,n\}},d_{\{\lambda,n\}}, f_{\{l,n\}}\right\}$. 
The transverse gauge field $A_\mu$ is expanded as 
\ben\nonumber
&&A_m(x,\tau)=\frac{in }{\tilde \lambda^2 } 
e_{\{\lambda,n\neq 0\}}\p_m \Phi_{\{\lambda,n\}}+d_{\{\lambda,n\}}\epsilon_{mn}\p^n\Phi_{\{\lambda,n\}}
+f_{\{l,n\}}\p_m\Psi_i,  \\ 
&&A_\tau(x,\tau)=e_{\{\lambda,0\}}\Phi_{\{\lambda,0\}}
+ e_{\{\lambda,n\neq 0\}}\Phi_{\{\lambda,n\neq 0\}}. 
\een
Substituting these modes in the action of the partition function (\ref{pi}), we obtain
\ben
\frac{\kappa}{4\pi}\int d^3x \varepsilon^{\mu\nu\rho}A_\mu\p_\nu A_\rho
&=&\kappa\Big\{
\tilde\lambda^2d_{\{\lambda,0\}}e_{\{\lambda,0\}} +  
\sum_{n>0}^\infty
\left(\tilde\lambda^2+n^2 \right)(e_{\{\lambda,n\}}d_{\{\lambda,-n\}}
-e_{\{\lambda,-n\}}d_{\{\lambda,n\}})\nn
\\&&+\sum_{n,l> 0}n(f_{\{l,-n\}}f_{\{-l,n\}}-f_{\{l,n\}}f_{\{-l,-n\}})\Big\}. 
\een
In the above equation we have consider the modes with fixed $\lambda$. 
In fact there is 
an integral over $\lambda$. Each mode in $\lambda$ occurs with a density of states 
given in (\ref{density}). 
Integrating out 
$\left\{e_{\{\lambda,n \}},d_{\{\lambda,n\}} ,f_{\{l,n\}}\right\}$
we obtain the partition function. 
As before note that all modes with $n\neq 0$  are complex and we use the relation in (\ref{conjugate}). 
and the modes  $e_{\{\lambda, 0 \}}, d_{\{\lambda, 0 \}}$ are real. 
Performing the gaussian integrations we obtain 
\begin{eqnarray} \nonumber
\log { \hat Z}  = - \int_0^\infty d\lambda \mu(\lambda) \left[ \log ( \kappa \tilde\lambda^2) 
+ \sum_{n=1}^\infty \log \left( \kappa^2 ( n^2 +   \tilde\lambda^2 
)^2 \right)   \right]  - \sum_{n, l =1}^\infty \log ( \kappa^2 n^2) . 
 \\
\end{eqnarray}
Here $\hat Z$ is defined by $Z=(\text{det}\,\Box)\,\hat Z$.
After some rearrangements we obtain
\begin{eqnarray}
\log { \hat Z}   &=&  - \int d\lambda \mu(\lambda)  \left( 
\log \kappa  + \sum_{n=1}^\infty  \log ( \kappa^2) \right)     -\sum_{l, n=1}^\infty \log(\kappa^2)
\\ \nonumber
& & -\int d\lambda \mu(\lambda) \left(  \log \tilde \lambda^2 
+ 2  \sum_{n=1}^\infty ( n^2 + \tilde \lambda^2)  
  \right)   - \sum_{n, l =1}^\infty  \log (n^2) .  
 \end{eqnarray}
 We use the $\zeta$ function regularization  to perform the sums involving 
 \begin{equation}
 \sum_{n=1}^\infty 1 =  -\frac{1}{2}. 
 \end{equation}
 This results in 
 \begin{eqnarray}
 \log { \hat  Z} &=& = - \frac{1}{2} \log \kappa 
   -\int d\lambda \mu(\lambda) \left(
 \sum_{n = -\infty}^\infty \log ( \tilde \lambda^2 + n^2)    \right),  
 \end{eqnarray}
where we have ignored the $\kappa$ and $L$ independent constant. 
We can now substitute  the partition function ${\cal \hat Z}$  in (\ref{pi}) 
and use the expression (\ref{jacob}) for the determinant of the massless scalar in $AdS_2$ to obtain 
\begin{eqnarray}
\log {Z} &=&  - \frac{1}{2} \log \kappa . 
\end{eqnarray}
%
Thus we obtain the result the the partition function of the Abelian Chern-Simons theory
on $AdS_2\times S^1$ independent of $L$ and 
depends on the coupling constant $\kappa$ precisely 
the same way as that of the Abelian Chern-Simons theory on the space $A_q$.  
It is interesting to note that the  discrete modes of the vector on $AdS_2$ played the crucial 
role in obtaining the dependence on $\kappa$. 
This resulted in the  expected relation between the parition function on the two conformally related spaces. 
It is also important to observe that the space of functions over which we performed the 
path integral were all normalizable functions on $AdS_2\times S^1$.

\section{Supersymmetry on AdS$_2\times$S$^1$}

We would like to consider a non-abelian Chern-Simon theory based  on a gauge group $G$ 
on $AdS_2\times S^1$ and evaluate its partition function. 
To do this we use the fact that the  field content of 
${\cal N}=2$ supersymmetric Chern-Simons theory, 
apart from the gauge field  of interest consists of only auxillary fields
without kinetic energy terms. 
Thus the ${\cal N}=2$ supersymmetric Chern-Simons theory  is equivalent 
to the  bosonic Chern-Simons theory with the same gauge group $G$. 
Therefore we can determine the partition function of the bosonic Chern-Simons theory 
on $AdS_2\times S^1$ by evaluating the partition function of 
the ${\cal N}=2$ supersymmetric Chern-Simons theory on 
$AdS_2 \times S^1$ using the technique of localization. 
In this section we first solve for  the 
Killing spinors on $AdS_2\times S^1$. We use these Killing spinors to construct the 
supersymmetric variation  of the ${\cal N}=2$ gauge multiplet and demonstrate the 
invariance of the Chern-Simons action. 
Finally we determine the locallizing term which is exact under the supersymmetric 
variation. 

\subsection{Killing spinors on $AdS_2\times S^1$}

The background metric  of $AdS_2 \times S^1$ is given by
\be\label{metric}
ds^2=d\tau^2+L^2(dr^2+\sinh^2r \,d\theta^2)\,.
\ee
The vielbein are $e^1=d\tau,\, e^2=L \,dr,\, e^3=L\sinh r \,d\theta$.\\ 
The non vanishing components of Christoffel symbols and spin connections are given by
\be
\Gamma^r_{\theta\theta}=-\cosh r\sinh r,\quad \Gamma^\theta_{r\theta}=\coth r,\quad \omega^3{}_2=\cosh r\, d\theta\,.
\ee\,
The Killing spinors are solutions of the following equations
\ben\label{killingspinor}
\l(\nabla_\mu-iA_\mu\r)\epsilon=-\frac{1}{2}H\gamma_\mu\epsilon-i V_\mu\epsilon-\frac{1}{2}\epsilon_{\mu\nu\rho}V^\nu\gamma^\rho\epsilon,  \nn\\
\l(\nabla_\mu+iA_\mu\r)\tilde\epsilon=-\frac{1}{2}H\gamma_\mu\tilde\epsilon+i V_\mu\tilde\epsilon+\frac{1}{2}\epsilon_{\mu\nu\rho}V^\nu\gamma^\rho\tilde\epsilon. 
\een\,
Here $A_\mu$, $H$ and $V_\mu$ are fields in the supergravity multiplet. At the linearised level $A_\mu$ couples to R-symmetry current, $H$ couples to string current and $V_\mu$ is the dual of graviphoton field strength which couples to central charge current. $\epsilon$ and $\tilde\epsilon$ are complex spinors parameterizing the supergravity transformations with R-charge $+1$ and $-1$, respectively. In Lorentzian signature, $\epsilon$ and $\tilde\epsilon$ are complex conjugate to each other but in Euclidean theory they are independent complex spinors.\\
It is clear from killing spinor equations (\ref{killingspinor}) that the vector field $K^\mu=\tilde\epsilon\gamma^\mu\epsilon$ is a Killing vector i.e. it satisfies the Killing vector equation
\be
\nabla_\mu K_\nu+\nabla_\nu K_\mu=0\,.
\ee
For our metric background the compact isometries are generated by killing vectors of the form $K_\pm=\frac{\p}{\p \tau}\pm\frac{\p}{\p\theta}$. We make the following choice of killing vector
\be
K=\frac{\p}{\p \tau}+\frac{1}{L}\frac{\p}{\p\theta}
\ee
This choice of Killing vector simplifies the Killing spinor and supergravity background fields which are given by
\ben\label{susy-backgrd}
&&\epsilon= e^{\frac{i\theta}{2}}\begin{pmatrix}i\cosh(\frac{r}{2})\\\sinh(\frac{r}{2})\end{pmatrix}\,,\qquad \tilde\epsilon=e^{-\frac{i\theta}{2}}\begin{pmatrix}\sinh(\frac{r}{2})\\i\cosh(\frac{r}{2})\end{pmatrix}\,,\nn\\ &&A_\tau=V_\tau=\frac{1}{L}\,,\quad A_{r,\theta}=H=0\,.
\een
For more details on the solutions of Killing spinor equations see the appendix B.
\subsection{Supersymmetry of  the vector multiplet}
Vector multiplet in $\mathcal N=2$ theory in Lorentzian signature contains a real scalar $\sigma$, gauge field $A_\mu$, an auxiliary real field $G$ and 2 component Weyl fermions $\lambda$ and $\tilde\lambda$. In order to compute partition function we need to analytically continue to Euclidean space. We choose the analytic continuation where the scalar field $\sigma$ and the auxiliary field $G$ are purely imaginary, the gauge field $A_\mu$ is real and the spinors $\lambda$ and $\tilde\lambda$ are two independent complex spinor.  
As we will see this choice of analytic continuation makes the bosonic part of the $Q$-deformation in the action positive definite.\\
The Euclidean supersymmetry transformation of the fields in a vector multiplet is given by
\ben
&&Q\lambda=-\frac{i}{4}\epsilon \,G-\frac{i}{2}\epsilon^{\mu\nu\rho}\gamma_\rho F_{\mu\nu}\epsilon-i\gamma^\mu\epsilon\l(iD_\mu\sigma-V_\mu\sigma\r)\,,\nn\\
&&Q\tilde\lambda=\frac{i}{4}\tilde\epsilon \,G-\frac{i}{2}\epsilon^{\mu\nu\rho}\gamma_\rho F_{\mu\nu}\tilde\epsilon+i\gamma^\mu\tilde\epsilon\l(iD_\mu\sigma+V_\mu\sigma\r)\,,\nn\\
&&QA_\mu=\frac{1}{2}\l(\epsilon\gamma_\mu\tilde\lambda+\tilde\epsilon\gamma_\mu\lambda\r)\,,\nn\\
&&Q\sigma=\frac{1}{2}\l(-\epsilon\tilde\lambda+\tilde\epsilon\lambda\r)\,,\nn\\
&&QG=-2i\l[D_\mu\l(\epsilon\gamma^\mu\tilde\lambda-\tilde\epsilon\gamma^\mu\lambda\r)-i\l[\sigma,\epsilon\tilde\lambda+\tilde\epsilon\lambda\r]-iV_\mu\l(\epsilon\gamma^\mu\tilde\lambda+\tilde\epsilon\gamma^\mu\lambda\r)\r]\,.
\een
The square of the susy transformations on vector multiplet fields are given by
\ben
&&Q^2\lambda=\mathcal L_K\lambda+i[\Lambda,\lambda]-\frac{1}{2L}\lambda\,,\nn\\
&&Q^2\tilde\lambda=\mathcal L_K\tilde\lambda+i[\Lambda,\tilde\lambda]+\frac{1}{2L}\tilde\lambda\,,\nn\\
&&Q^2A_\mu=\mathcal L_K A_\mu+D_\mu\Lambda\,,\nn\\
&&Q^2\sigma=\mathcal L_K\sigma-iK^\mu[A_\mu,\sigma]\,,\nn\\
&&Q^2G=\mathcal L_K G+i[\Lambda,G]\,.
\een
Here $\Lambda=\tilde\epsilon\epsilon\,\sigma-K^\rho A_\rho$. \\Using the above supersymmetry transformations we also note that $Q\Lambda=0$.\\
Therefore the algebra of supersymmetry transformation is given by
\be
Q^2=\mathcal L_K+\delta^{\text{gauge transf}}_\Lambda+\delta^{R-\text{symm}}_{\frac{1}{2L}}\,.
\ee
It is equivalent to work with fermion bilinear $(\Psi,\Psi_\mu)$ instead of $(\lambda,\tilde\lambda)$ which are defined as 
\be
\Psi=\frac{i}{2}(\tilde\epsilon\lambda+\epsilon\tilde\lambda)\,,\quad \Psi_\mu=QA_\mu=\frac{1}{2}(\epsilon\gamma_\mu\tilde\lambda+\tilde\epsilon\gamma_\mu\lambda)\,.
\ee
The fermion bi-linears are convenient for the evaluation of the index. 
The inverse of the above relations expresses $(\lambda,\tilde\lambda)$ in terms of  $\Psi,\Psi_\mu$ as
\ben
\lambda=\frac{1}{\tilde\epsilon\epsilon}\l[\gamma^\mu\epsilon\Psi_\mu-i\epsilon\Psi\r],\quad \tilde\lambda=\frac{1}{\epsilon\tilde\epsilon}\l[\gamma^\mu\tilde\epsilon\Psi_\mu-i\tilde\epsilon\Psi\r]\,.
\een
The supersymmetry transformation of the bi-linears are
\ben
&&Q\Psi=\frac{1}{4}(\tilde\epsilon\epsilon)G-\frac{i}{2}\l(\tilde\epsilon\gamma^{\mu\nu}\epsilon\r) F_{\mu\nu}-\frac{1}{L}\sigma\,,\nn\\
&&Q\Psi_\mu=\mathcal L_K A_\mu+D_\mu\Lambda\,.
\een

\subsection{The localizing  action}

Next we deform the action by a $Q$-exact term, $t\,QV_{{\rm loc} }$. According to the principle of supersymmetric localization, the partition function does not depend on the parameter $t$ and the choice of $V_{{\rm loc} }$. Thus one can take $t$ to infinity. In this limit the path integral receives contribution from the field configurations which are minima of $QV_{{\rm loc} }$. 
One convenient choice of $V_{{\rm loc}}$ is given by
\be
V_{{\rm loc} }=\int d^3x\sqrt{g}\frac{1}{(\tilde\epsilon\epsilon)^2}\text{Tr}\l[\Psi^\mu (Q\Psi_\mu)^\dagger+\Psi (Q\Psi)^\dagger\r]\,.
\ee
The bosonic part of the $QV_{{\rm loc} }$ action is given by
\ben\label{bosonLagr}
& & QV_{{\rm loc} \{\text{bosonic}\}}=\int d^3x\sqrt{g}\frac{1}{2(\tilde\epsilon\epsilon)^2}\text{Tr}\l[(Q\Psi^\mu)(Q\Psi_\mu)^\dagger+(Q\Psi)(Q\Psi)^\dagger\r]\nn\\&=&\int d^3x\sqrt{g}\text{Tr}\l[\frac{1}{4}F_{\mu\nu}F^{\mu\nu}-\frac{1}{2\cosh^2r}D_\mu(\cosh r\,\sigma)D^\mu(\cosh r\,\sigma)-\frac{1}{32}\l(G-\frac{4\sigma}{L\cosh r}\r)^2\r]\,.\nn\\
\een
The minima of  $QV_{{\rm loc} \{\text{bosonic}\}}$ are the solutions of the following equations
\be
F_{\mu\nu}=0\,,\quad D_\mu(\cosh r\,\sigma)=0\,,\quad G=\frac{4\sigma}{L\cosh r}\,.
\ee 
In order to solve $F_{\mu\nu}=0$ for the background gauge field $A_\mu$, we choose a gauge $A_r=0$. In this gauge we get the following equations for $A_\mu$\,, 
\begin{eqnarray}
&& F_{\tau r}=0\Rightarrow \p_r A_\tau=0\,,\\
&& F_{r\theta}=0\Rightarrow \p_r A_\theta=0\,,\\
&& F_{\tau \theta}=0\Rightarrow \p_\tau A_\theta-\p_\theta A_\tau -i\l[A_\tau,A_\theta\r]\,.
\end{eqnarray}
The first two equations imply that both $A_t$ and $A_\theta$ are independent of $r$. Requiring that the solutions should be normalizable imply that $A_t=0$. The third equation imply that $A_\theta$ is independent of $t$ and thus can only be function of $\theta$.
Now requiring further that the solution of localization equation should be smooth near the origin of AdS$_2$ implies that $A_\theta$ should approach to zero. Thus the only solution for the localization equation $F_{\mu\nu}=0$ is trivial and there are no non trivial smooth and normalizable solutions for $A_\mu$.
Thus the solution of localization equation upto gauge transformations is given by
\be
A_\mu=0\,,\quad\sigma=\frac{i\alpha}{\cosh r}\,,\quad G=\frac{4i\alpha}{L\cosh^2r}
\ee
Here $\alpha$ is a real constant matrix valued in Lie algebra. Furthermore on this localization background the gauge transformation parameter in supersymmetry algebra reduces to a constant, $\Lambda^{(0)}=i\alpha$.\\
The supersymmetric completion of a bosonic Chern-Simons action is given by
\be\label{SCSaction}
S_{\text{C.S.}}=\int d^3x\sqrt{g}\text{Tr}\l[i\varepsilon^{\mu\nu\rho}\l(A_\mu\p_\nu A_\rho-\frac{2i}{3}A_\mu A_\nu A_\rho\r)-\tilde\lambda\lambda+\frac{i}{2}G\sigma\r]\,.
\ee
Here $\varepsilon^{\mu\nu\rho}=\frac{1}{\sqrt{g}}\epsilon^{\mu\nu\rho},\quad \epsilon^{\tau\eta\theta}=1$.\\
We note here that the fermions and scalars in the vector multiplet are purely auxiliary fields as they do not have kinetic terms and therefore, one can integrate them out. Thus the supersymmetric Chern-Simons theory is equivalent to a bosonic Chern-Simons theory.\\ 
The action in (\ref{SCSaction}) is invariant under supersymmetry transformation upto terms which are total derivative 
\be
QS_{\text{C.S.}}=\int d^3x\sqrt{g}\nabla_\rho\text{Tr}\l[\frac{i}{2}\varepsilon^{\rho\mu\nu}A_{\mu}(\epsilon\gamma_{\nu}\tilde\lambda+\tilde\epsilon\gamma_{\nu}\lambda)+(\epsilon\gamma^\rho\tilde\lambda-\tilde\epsilon\gamma^\rho\lambda)\sigma\r]\,.
\ee
In terms of cohomological variable the above boundary terms can be written as
\be\label{SusyVar.CS}
QS_{\text{C.S.}}=\int d^3x\sqrt{g}\nabla_\rho\text{Tr}\l[\frac{i}{2}\varepsilon^{\rho\mu\nu}A_{\mu}\Psi_\nu+\frac{2i}{\tilde\epsilon\epsilon}(K^\rho\Psi+\varepsilon^{\rho\mu\nu}K_\mu\Psi_\nu)\sigma\r].
\ee
We will comment more about the boundary terms later when we discuss about the boundary conditions on fields and show that there are no contributions from boundary terms.
\section{The one loop determinant}
We will now proceed to compute the determinant coming from the quadratic fluctuation of 
$QV_{{\rm loc} }$ action. 
This is done  by  obtaining 
 the indices of the operators involved in the one loop 
determinant by explicitly solving 
for the solutions and counting the ones which contribute to the index. 
To begin, 
we  simplify our path integral using the gauge invariance which allow us to diagonalize the 
Lie algebra valued matrix $\alpha$ of the gauge  group $G$. 
This introduces the Vandermonde determinant in the path integral. 
Thus our path integral becomes
\be\label{partition.fn.}
Z=\int d\alpha \,\prod_{\rho>0}(\rho.\alpha)^2\,\exp\l(\frac{\kappa}{{4\pi}}S_{\text{C.S.}}\r) Z_{1-\text{loop}}(\alpha)\,.
\ee
We will now compute $Z_{1-\text{loop}}(\alpha)$. 
\subsection{Localization in $U(1)$ Chern Simons theory}

Let us again begin with the warm up example  of the one loop determinant 
about the localizing solution for the case of  the Abelian theory for which the
evalution  of the one loop determinant 
simplifies considerably.   In this case it is very 
easy to see that the one loop determinant is trivial i.e. independent of the parameter $\alpha$. 
The bosonic part of the action  given in  equation (\ref{bosonLagr}) at the quadratic order in 
fluctuations do not have any dependence on $\alpha$. Therefore, the one loop 
determinant coming from bosonic fluctuations will not have $\alpha$ dependence. 
The fermionic part of the $QV_{{\rm loc} }$-Lagrangian in the Abelian case also 
does not depend on $\alpha$. The fermionic part of the Lagrangian is given by 
\ben
QV_{{\rm loc}\{\text{fermionic}\}}=
\text{Tr}\Big[-\frac{i}{2L(\tilde\epsilon\epsilon)}(\lambda\tilde\lambda)
-\frac{i}{2}V^a(\tilde\lambda\gamma_a\lambda)
-\frac{1}{2}\tilde\lambda\slashed{D}\lambda
-\frac{1}{2}\lambda\slashed{D}\tilde\lambda\nn\\
+\frac{i}{4}\frac{V^a(\epsilon\gamma_a\epsilon)}{\tilde\epsilon\epsilon}(\tilde\lambda\tilde\lambda)
+\frac{i}{4}\frac{V^a(\tilde\epsilon\gamma_a\tilde\epsilon)}{\tilde\epsilon\epsilon}(\lambda\lambda)\Big]. 
\een
Therefore, the one loop determinant coming from fermionic fluctuations will also not have $\alpha$ dependence. Thus upto a normalization constant, the partition function is completely determined by the classical action. \\
The classical action evaluated at the localizing solution is given by
\be
\exp\l(\frac{\kappa}{{4\pi}}S_{\text{C.S.}}\r)=\exp(-{\pi}i\alpha^2L\kappa). 
\ee
Thus the partition function of the Abelian Chern Simons theory is given by
\be\label{AbelianCSint.}
Z\sim \int d\alpha \exp(-{\pi}i\alpha^2L\kappa)\sim \frac{1}{\sqrt{\kappa L}}\,.
\ee
As we will show later, a more careful analysis of the determinant shows that the normalization constant do depends on $L$ and in particular it exactly cancel the $\sqrt{L}$ coming from the integral.
\subsection{Localization in Non-abelian Chern Simons theory}
We will now compute the one loop determinant about the localizing solution for non abelian Chern Simons gauge theories. For this, first we need to introduce the gauge fixing Lagrangian. This could be achieved by choosing any convenient gauge condition. In our case it turns out that the analysis becomes simpler for the gauge fixing Lagrangian\,\footnote{As we will show in the subsection 4.3, this gauge choice fixes the gauge completely. It is also possible to show that this gauge choice intersects every gauge orbit.}
\be\label{gfLagrangian}
\mathcal L_{g.f.}=\text{Tr}\,Q_B\l[i\tilde c\nabla_\mu\l(\frac{1}{\cosh^2r}a^\mu\r)+\frac{1}{2}\xi \tilde c B\r]\,.
\ee
As we will show below the complete action including the gauge fixing Lagrangian is invariant under BRST transformations on the fields which are given by
\ben
&&Q_Ba_\mu=D_\mu c,\quad Q_B\tilde c=B,\quad Q_B c=\frac{i}{2}\{c,c\},\quad Q_B\tilde\lambda=i\{c,\tilde\lambda\}\nn\\&&Q_B\lambda=i\{c,\lambda\},\quad Q_B\hat\sigma=i[c,\hat\sigma],\quad Q_B\hat G=i[c,\hat G],\quad Q_BB=0\,.
\een
Here $a_\mu,\,\hat\sigma$ and $\hat G$ are fluctuations away from localizing . \\
We also define the susy transformations for extra fields
\be
Qc=-\Lambda+\Lambda^{(0)},\quad QB=\mathcal L_K\tilde c+i[\Lambda^{(0)},\tilde c],\quad Q\tilde c=0
\ee
such that the combined transformations generated by $\hat Q=Q+Q_B$ satisfy the algebra
\be\label{Qhatalgebra}
\hat Q^2=\mathcal L_K+\delta^{\text{gauge transf.}}_{\Lambda^{(0)}}\,.
\ee
To summarize, the complete transformations of fields under $\hat Q$ are given by
\ben\label{QhatTransf}
&&\hat Qa_\mu=\Psi_\mu+D_\mu c,\qquad \hat Q\hat\sigma=Q\hat\sigma+i[c,\hat\sigma]\,,\nn\\
&&\hat Q\Psi_\mu=\mathcal L_K a_\mu+D_\mu\Lambda+i\{c,\Psi_\mu\},\quad \hat Q\Psi=
\frac{1}{4}(\tilde\epsilon\epsilon)\hat G
-\frac{i}{2}\l(\tilde\epsilon\gamma^{\mu\nu}\epsilon\r) F_{\mu\nu}(a)-\frac{1}{L}\hat\sigma+i\{c,\Psi\}\,,\nn\\
&&\hat Qc=-\Lambda+\Lambda^{(0)}+\frac{i}{2}\{c,c\},\qquad \hat Q\tilde c=B\,.
\een
The gauge fixing Lagrangian (\ref{gfLagrangian}) is not $Q$ closed and therefore we can not use it for 
the localization problem. But a simple modification of it allows us to use 
\ben
\hat{\mathcal L}_{g.f.}&=&\text{Tr}\,\hat Q\l[i\tilde c\nabla_\mu\l(\frac{1}{\cosh^2r}a^\mu\r)
+\frac{1}{2}\xi \tilde c B\r], \nn\\&=&{\mathcal L}_{g.f.}
-i\text{Tr}\,\tilde c\nabla^\mu\l(\frac{1}{\cosh^2r}\Psi_\mu\r)
-\frac{1}{2}\text{Tr}\,\xi\tilde c\l(\mathcal L_k\tilde c+i[\Lambda^{(0)},\tilde c]\r). 
\een 
Clearly $\hat{\mathcal L}_{g.f.}$ is equivalent to ${\mathcal L}_{g.f.}$ as the rest of the terms do not 
contribute to the partition function. With $\hat{\mathcal L}_{g.f.}$ our path integral is 
invariant under $\hat Q$ and we will use $\hat Q$ for the localization of the path integral. 

\subsubsection*{Boundary  conditions}

When   evaluating  
partition functions on  non compact spaces it is important to specify the choice 
of boundary conditions obeyed by the fields in the theory. 
The partition function  depends on  the boundary conditions chosen.  
Furthermore to apply  the method of 
supersymmetric localization 
 the boundary conditions must also be invariant under the  supersymmetry transformations. 
 In order to achieve this, we begin by 
 imposing the boundary conditions on the fields 
 $(A_\mu,\Psi,c ,\tilde c)$ and then impose the boundary conditions on 
 the rest of the fields following supersymmetry transformations. For example we require that 
 the field $\hat QA_\mu$ has the same boundary condition as that of $A_\mu$.\\
Looking at the $\hat QV$ Lagrangian we find that in order for the vector field to be square 
integrable, the components must satisfy the following asymptotic conditions as $r\rightarrow \infty$ 
\be\label{bdy.cond1}
e^{r/2}A_{\tau} \rightarrow 0 ,\,\,e^{r/2}A_{r} \rightarrow 0, \,\, e^{-r/2}A_\theta \rightarrow 0. \,
\ee
The boundary condition on the ghost field $c$ is determined by using the fact that it is a 
gauge transformation parameter. It is required that the gauge transformations should 
not change the asymptotic boundary condition of  the  gauge field given in (\ref{bdy.cond1}. 
This forces the 
ghost field $c$ to have the following asymptotic behaviour as $r\rightarrow\infty$ 
\be
c \sim f(\theta)+\tilde f(\theta,\tau)\,e^{-r/2}+...\,.
\ee
Once we fix the asymptotic behaviour of the allowed gauge transformations, the Faddeev-Popov determinant needs to be computed in this restricted subspace of the gauge transformations. This naturally fixes the $\tilde c$ to have the same boundary condition as that of $c$\,\footnote{The ghost Lagrangian density in our case is given by $\tilde c\nabla_
\mu(\frac{g^{\mu\nu}}{\cosh^2r}\nabla_\nu c)$. Defining $c=\cosh r\,\omega_1$ and $\tilde c=\cosh r\,\omega_2$, where $\omega_{1,2}$ are scalar fields, we see that the corresponding operator for $\omega_{1,2}$ is a self adjoint and maps a mode with given asymptotic behaviour to a mode with the same asymptotic behaviour. }. \\The boundary condition on the field $\Psi$ is determined by the asymptotic behaviour of its $\hat Q$ variation (\ref{QhatTransf}). Given the asymptotic behaviour of the gauge field $A_\mu$ and $G$\,\footnote{From the bosonic part of $\hat QV$ Lagrangian, we see that the field $G$ is square integrable if $e^{r/2}G\rightarrow 0$ as $r\rightarrow \infty$.}, we see that $\hat Q\Psi$ satisfies the condition $e^{-r/2}\hat Q\Psi\rightarrow 0$ as $r\rightarrow \infty$. We, therefore, require that the field $\Psi$ satisfies the boundary condition $e^{-r/2}\Psi\rightarrow 0$.    \\ 
Next we will discuss the smoothness conditions for the fields near $r\rightarrow 0$ which play an 
important role in analysing the space of kernel and cokernel in the next sections. In order to find the 
regularity conditions near $r\rightarrow 0$, we expand the field in 
terms of Fourier mode, $X(\tau,r,\theta)=X^{(n,p)}(r)e^{(in\tau+ip\theta)}$. Near 
$r\rightarrow 0$, the $\theta$-circle shrinks to zero size and therefore, the 
regular behaviour of the field is determined by integer $p$. For any scalar field, collectively 
denoted by $\Phi$, its Fourier mode $\Phi^{(n,p)}$ needs to have $\sim r^p$ behaviour as $r\rightarrow 0$. On the other hand 
the component of a vector field should satisfy the following regularity conditions
\ben
&&A^{(n,p\neq 0)}_\tau\sim r^{p},\quad  A^{(n,p\neq 0)}_r\sim r^{p-1},\quad 
A^{(n,p\neq 0)}_\theta\sim r^{p}, \nn\\
&& A^{(n,p= 0)}_\tau\sim \mathcal O(1),\quad  A^{(n,p= 0)}_r\sim r,\quad A^{(n,p=0)}_\theta\sim r^{2}. 
\een
Now let us look at the variation of Chern Simons action under $\hat Q$. Compared
 to (\ref{SusyVar.CS}) we obtain extra terms proportional to the ghost field, 
\ben
\hat QS_{\text{C.S.}}=\int d^3x\sqrt{g}
\nabla_\rho\text{Tr}\l[\frac{i}{2}\varepsilon^{\rho\mu\nu}A_{\mu}\hat\Psi_\nu+\frac{2i}
{(\tilde\epsilon\epsilon)^2}\varepsilon^{\rho\mu\nu}K_\mu(\hat\Psi_\nu-D_\mu c)(\Lambda+K^\mu A_\mu)\r]\,,\nn\\
\een
where  $\hat\Psi_\mu=\hat QA_\mu=\Psi_\mu+D_\mu c$. With the above boundary conditions, both at $r\rightarrow 0$ and $r\rightarrow \infty$, we see that the $\hat Q$ variation of the supersymmetric Chern Simons action vanishes and thus we can use the techniques of  localization to compute the partition function.\\
To proceed further we change the field variables to $X_0  = (a_{\mu})$, $X_1 =(\Psi, c, \tilde{c})$ and $X_0'= \hat Q X_0$ and $X_1'= \hat Q X_1$. In this notation our set of bosonic fields $(a_\mu,\Lambda,\hat G,B)$ are represented by $(X_0,X_1')$ and fermionic fields $(\Psi,\Psi_\mu,c,\tilde c)$ are represented by $(X_1,X_0')$. We then rewrite our localization Lagrangian, if necessary integrating
by parts. It is important to note 
that because of the presence of $\frac{1}{\cosh^2r}$ 
and the asymptotic boundary conditions on all the fields, there are no boundary terms 
while integrating by parts. We thus obtain the localization Lagrangian 
\begin{align}
 V_{{\rm loc} }=\text{Tr}\,(\hat Q X_0~ & X_1)\left(
\begin{array}{cc}
 \text{D}_{00} & \text{D}_{01} \\
 \text{D}_{10} & \text{D}_{11} \\
\end{array}
\right)\left(
\begin{array}{c}
X_0 \\
 \hat Q X_1 \\
\end{array}
\right)\,.
\end{align}
Here $D_{ij}$ are various 
differential operators and we are only keeping the quadratic terms in the fluctuations. 
By taking $\hat Q$ of $ V_{loc}$ and 
assembling bosonic and fermionic terms one can show that the one-loop result is :
\begin{equation}
 Z_{1-\text{loop}}(\alpha)=\sqrt{\frac{{\rm Det}_{\rm{Coker} D_{10}}(\hat Q^2)}{{\rm Det}_{\rm{Ker}  D_{10}}(\hat Q^2)}}\,.
\end{equation}
Thus for a given eigen value of $H\equiv \hat Q^2$, we just need to know the difference in dimensions of kernel and cokernel of $D_{10}$ operator. This difference is encoded in the equivariant index of $D_{10}$ defined as
\be
\text{ind}_{\text{equiv}}D_{10}=\text{Tr}_{\text{ker}D_{10}}e^{iHt}-\text{Tr}_{\text{coker}D_{10}}e^{iHt}=\sum_{n\in \mathbb Z}(m^{(0)}_n-m_n^{(1)})q^n\,.
\ee
Here $m_n^{(0,1)}$ are the dimension of kernel and cokernel for a given eigen 
value of $H$ labelled by $n$.\\
To identify the $D_{10}$ operator, it is convenient to note the following: 
\begin{eqnarray}
&&(\hat Q \Psi_{\mu})^{\dagger} = {\cal L}_K a_{\mu} +(D_{\mu} \Lambda)^{\dagger} +\dots =  {\cal L}_K a_{\mu} - D_{\mu} \Lambda -2 D_{\mu}(K.a)+\dots, \\
&&(\hat Q \Psi)^{\dagger}=-\hat Q \Psi- i(\tilde{\epsilon}\gamma^{\mu \nu} \epsilon) F_{\mu \nu}+\dots
\end{eqnarray}
where dots  include terms quadratic in fermionic fields. Since we are only interested in $D_{10}$ the relevant terms are 
\begin{eqnarray}\label{D10equ}
 -\frac{\sqrt{g}}{(\tilde\epsilon\epsilon)^2} \text{Tr}\,\Big[g^{\mu\nu}D_{\mu} c\, ( {\cal L}_K a_{\nu}+[\alpha,a_
 \nu]-2 D_{\nu}(K.a))+ i\Psi (\tilde{\epsilon} \gamma^{\mu \nu} \epsilon) F_{\mu \nu}
 -i\tilde{c}(\tilde\epsilon\epsilon)^2 D^{\mu} \l( \frac{1}{(\tilde\epsilon\epsilon)^2}  a_{\mu} \r) \Big]\nn\\
\end{eqnarray} 
Before proceeding for the detailed computations of kernel and cokernel, we notice that the pure 
constant mode of $c$ and $\tilde c$ are zero modes of the $\hat QV$ action. We, therefore, consider the path integral measure such that we don't integrate over the constant zero modes.\\
Since $\hat Q^2$ commutes with $\partial_t$ and $\partial_{\theta}$, we can study the kernel and 
cokernel for each Fourier mode 
in $t$ and $\theta$ separately. We, therefore, write $X_0$ fields as $X_0(r) e^{-i(nt+p\theta)}$ and $X_1$ fields as $X_1(r) e^{i(nt+p\theta)}$. Note that the eigenvalues of $\hat Q^2$ on $X_{0,1}$ in this subspace are $-i(n+\frac{p}{L}\pm i \rho\cdot\alpha))$ where $\rho$ is a root vector.\\
\subsection{Analysis for kernel and co-kernel}\label{Analysis_of_Equ}
It is convenient to make the following change of variables in each root space $\rho$ :
\begin{eqnarray}\label{newvariable}
&&a_t= a(K) - a_{\theta}/L,\\
&&\tilde c=\hat{\tilde c}+\frac{Ln+p-L\rho\cdot\alpha}{L}c ,
\end{eqnarray}
where $a(K)= K\cdot a$. Note that the boundary condition of $\hat{\tilde c}$ is the same as $\tilde c$ and that of $a(K)$ is the same as
$a_{\theta}$.

By varying (\ref{D10equ}) with respect to $c$, $\hat{\tilde{c}}$ and $\Psi$ we get the following equations for the kernel:
\begin{eqnarray}\label{kernelequ.}
 \Delta^{(n,p)} a(K)(r)=0\,,\nn\\
 i \partial_r \l(\frac{\sinh r}{\cosh^2r}a_r(r)\r)+ \l(\frac{p}{\sinh^2 r} - 
Ln\r)\frac{\sinh r}{\cosh^2r}a_{\theta}(r)+\frac{\sinh r}{\cosh^2r}L^2 n a(K)=0\,, \nn \\ 
\partial_r a_{\theta}(r)+  
i \l(\frac{p}{\sinh^2r} - Ln \r)\frac{\sinh^2r}{\cosh^2r} a_r(r)-L \tanh^2 r\partial_r a(K)(r)=0\,.
\end{eqnarray}
and by varying $a(K)$, $a_r$ and $a_{\theta}$ the following equations for the co-kernel:
\begin{eqnarray}\label{cokernelequ.}
\Delta^{(n,p)} c(r)+\frac{1}{2 L \sinh r}\partial_r \l(\tanh^2 r\Psi \r)+\frac{n }{4 \cosh^2 r}\hat{\tilde{c}}=0\,,\nn\\
 -2\partial_r \Psi(r)+  \l(\frac{p}{\sinh^2r} - 
L n\r)\frac{\sinh r}{\cosh^2r}\hat{c}(r)=0\,, \nn \\ 
\frac{\sinh r}{\cosh^2r}\partial_r\hat{c}(r)- 2
\l(\frac{p}{\sinh^2r} - Ln \r)\tanh^2 r\,\Psi(r)=0\,.
\end{eqnarray}
where $\Delta^{(n,p)} =\nabla^{\mu} \frac{1}{|\epsilon|^2} \nabla_{\mu}$ is a second order differential operator restricted to 
Fourier mode labelled by $(n,p)$. Note that these equations do not have any $\alpha$-dependence and 
therefore they are the same in all the root spaces.  We make a few remarks on the above equations: \\

\noindent 1) The first kernel equation is a decoupled equation for $a(K)$ involving the operator $\Delta^{(n,p)}$. 
We will show below that it has no smooth solution satisfying the asymptotic boundary
condition, except for a constant solution for $(n,p)=(0,0)$ . This means that $a(K)$ decouples from the last two kernel equations 
and they become coupled homogeneous first order differential 
equations for $a_r$ and $a_{\theta}$. It is worth noting that the gauge transformation by some function $f(r)e^{i(n t+ p \theta)}$
changes the gauge-fixing condition 
$\nabla^{\mu}(\frac{1}{\cosh^2 r} A_{\mu})$ by $\Delta^{(n,p)}f(r)$. The fact that $\Delta^{(n,p)}f(r)=0$ has no non-trivial solutions in the 
allowed space of gauge transformations $f(r)$ \footnote{ $(n,p)=(0,0)$ and constant $f$ defines gauge transformation by constant which doesn't change the gauge 
fields}, shows that our gauge fixing condition indeed fixes the gauge completely.\\

\noindent 2) The homogeneous part of the first co-kernel equation for $c$ therefore has also no solution. 
It may however have particular solution due to the inhomogeneous terms.\\

\noindent 3) The last two co-kernel equations are first order differential equations in $\hat{\tilde{c}}$ and $\Psi$ and do not involve $ c$. 
These equations are identical to the last two kernel equations (after setting $a(K)=0$) with the identification
$a_{\theta} \rightarrow \hat{\tilde{c}}$ and $a_r \rightarrow  2 i \frac{\cosh^2r}{\sinh r}\Psi$ \\

\noindent {\bf Kernel Equations}\\

First we show that $\Delta^{(n,p)} a(K)(r) =0$ has no allowed solution for $(n,p) \neq (0,0)$. The following argument will also show that 
for $(n,p)=(0,0)$ there is one solution $a(K)(r)=\rm{constant}$.  
The explicit form of this equation is given by 
\ben
-\sinh^2r\,\p^2_ra(K)(r)-\tanh r(1-\sinh^2r)\,\p_ra(K)(r)+(p^2+n^2L^2\sinh^2r)a(K)(r)=0\,.\nn\\
\een
This equation has real coefficients, therefore without losing generality, we can assume that the solution $a(K)(r)$ is real in $r \in [0,\infty]$. Multiplying 
this equation by $a(K)(r)/(\cosh^2r~\sinh r) $ we obtain:
\begin{eqnarray}
 L_k\equiv &~&\frac{\sinh r}{\cosh^2 r} \l( (\partial_r a(K)(r))^2 + (p^2+ L^2 n^2 \sinh^2 r) a(K)(r)^2 \r)\nonumber
\\ &-& \partial_r 
 \l(\frac{\sinh r}{\cosh^2 r} a(K)(r)\partial_r a(K)(r)\r)
\end{eqnarray}
If $a(K)$ satisfies the equation, then $L_k$ must be zero. Now integrating $L_k$ we obtain the condition
\begin{eqnarray}
 0=\int_0^{\infty} dr L_k = \int_0^{\infty}&dr&\frac{\sinh r}{\cosh^2 r} \l( (\partial_r a(K)(r))^2 + (p^2+ L^2 n^2 \sinh^2 r) a(K)(r)^2 \r)\nonumber\\
 &-& \l(\frac{\sinh r}{\cosh^2 r} a(K)(r)\partial_r a(K)(r)\r)|_0^{\infty}
\end{eqnarray}
The boundary term vanishes at $r=0$ for smooth $a(K)$ and vanishes at $r=\infty$ for $a(K)$ satisfying the asymptotic 
behaviour $a(K) e^{-r/2} \rightarrow 0$. The integrand on the right hand side is non-negative for all $r$, therefore for $(n,p) \neq (0,0)$, $a(K)$ must vanish. 
On the other hand for $(n,p)=(0,0)$, the above condition implies that $\partial_r a(K)(r)=0$ which allows for a constant solution $a(K)(r)= C_1$. 

In either case
$a(K)$ disappears from the remaining two kernel equations: in the third kernel equation only $\partial_r a(K)$ appears, while in the second  kernel equations $a(K)$ 
together with a factor of $n$. It is clear from the structure of the 
equations that the solutions can be assumed to be such  that $a_{\theta}$  is real and $a_r$ is pure imaginary (or vice versa). 
Multiplying the second kernel equation 
by $(- i \sech r~ \tanh r~ a_r(r))$ and the third kernel equation by $(\sech^2 r~ a_{\theta}(r))$ and adding them together one obtains: 
\begin{equation}
 S_k\equiv \frac{\tanh r }{\cosh^2r} a_{\theta}(r)^2 + \partial_r \l(\frac{1}{2 \cosh^2 r}( a_{\theta}(r)^2 +\tanh^2 r~ a_{r}(r)^2 \r)
\end{equation}
As $S_k$ is a linear combination of the two equations, $S_k$ will be zero on the kernel. 
The first term in $S_k$ is non-negative for all values of $r$ and the second term is a total derivative in $r$. 
This means that if we integrate $S_k$ over $r$ from 
$0$ to  $\infty$
\begin{equation}
0= \int_0^{\infty} dr S_k = \int_0^{\infty} dr\frac{\tanh r }{\cosh^2r} a_{\theta}(r)^2 + 
 \l(\frac{1}{2 \cosh^2 r}( a_{\theta}(r)^2 +\tanh^2~ r a_{r}(r)^2 )\r)|_0^{\infty}
\label{actionintegralK}
\end{equation}

From the smoothness condition at $r=0$, namely, for $p\neq 0$, $( a_r,a_{\theta}) \rightarrow (r^{|p|-1},r^{|p|})$ and for $p=0$,  
$( a_r,a_{\theta}) 
\rightarrow  (r,r^2)$,
we see that the boundary term in $S_k$, at $r=0$, vanishes for any smooth configuration. The boundary term at $r=\infty$ vanishes 
for square integrable gauge fields. 
This shows that $a_{\theta}=0$  and from the second kernel equation, it follows, that there is no solution for $a_r$ which is square integrable. Thus,
$a_r$ and $a_{\theta}$ vanish for all $n$ and $p$. Now $a_t= a(K) - a_{\theta}/L= a(K)$. For $(n,p)\neq (0,0)$, we have already shown that $a(K)=0$. For 
$(n,p) = (0,0)$, there was one solution  $a(K)=C_1$. This implies that for $(n,p) = (0,0)$,  $a_t = C_1$. However this  is not square integrable and 
therefore $C_1=0$.
Thus we have shown that 
in the space of smooth and square-integrable gauge fields, kernel vanishes for all $n$ and $p$.
\\

\noindent {\bf Co-kernel equations}\\

As discussed above, the homogeneous part of the first co-kernel equation for $c(r)$ is the same as the kernel equation for $a(K)$ 
for which we already showed that 
there is no solution. However there may be a solution for the inhomogeneous equation. In order to find the form of the inhomogeneous 
terms we need to find the 
solutions for $ \hat{\tilde c}$ and $\Psi$ using the last two co-kernel equations.
Both the equations involve real coefficients and therefore the solutions for $\hat{\tilde c}$ and $\Psi$ can be chosen to be real. 
Multiplying the third equation by $2 \Psi$ and the second equation by   $\frac{\hat{\tilde c}}{\sinh r}$ and adding them , one gets:
\begin{equation}
S \equiv \frac{\sinh r }{\cosh^2r} \hat{\tilde c}^2 + \partial_r \l(\frac{1}{2 \cosh^2 r} \hat{\tilde c}^2 -2 \Psi^2 \r)
\end{equation}
The second term in $S$ is a total derivative while the first term is non-negative. If the equations are satisfied then $S$ must be 
zero. Integrating over $r$ one gets:
\begin{equation}
 0=\int_0^{\infty} dr S = \int_0^{\infty} dr \frac{\sinh r }{\cosh^2r} \hat{\tilde c}^2 + 
\l(\frac{1}{2 \cosh^2 r} \hat{\tilde c}^2 -2 \Psi^2 \r) |_0^{\infty}
\label{actionintegral}
\end{equation}
The boundary term at $r=0$ vanishes for $p \neq 0$ as both $\Psi$ and $\hat{\tilde c}$ must go as $r^{|p|}$. For $p=0$, however, 
the boundary term at $r=0$ 
may not be zero. Indeed one can study the series solutions of the two co-kernel equations for $\Psi$ and $\hat{\tilde c}$ and show 
that the solutions for $p=0$ 
go as $r^0$.

To analyse the boundary term at $r=\infty$, we look for the asymptotic solutions in a series expansion in $e^{-r}$. 
For $n \neq 0$, the asymptotic behaviours are 
$(  \hat{\tilde c}, \Psi) \rightarrow 
(e^{r\gamma_+ },e^{r(\gamma_+-1) })$ and 
$(  \hat{\tilde c}, \Psi) \rightarrow (e^{r\gamma_-},e^{r(\gamma_--1 ) })$ respectively, where
\begin{equation}\label{gamma}
\gamma_{\pm}=\frac{1}{2}(1 \pm \sqrt{1+ 4 L^2 n^2}). 
\end{equation}
Clearly only the second solution satisfies our 
boundary conditions, and 
for this the boundary term in $S$ vanishes at $r=\infty$. This proves that  there are no acceptable 
solutions to $(  \hat{\tilde c}, \Psi)$
for $n$ and $p$ both non-zero. This also implies that $c$ must be zero as there are no 
inhomogeneous terms in the first co-kernel equation. 

We already saw for $p=0$ with $n \neq 0$, that the boundary term in $S$ at $r=0$ does not vanish for smooth solution. 
The same happens for $n=0$ and $p \neq 0$ at $r=\infty$. The asymptotic behaviour of the two solutions,
for $n=0$ and $p \neq 0$, that can be verified by making a series expansion in $e^{-r}$, are 
\begin{eqnarray}
(  \hat{\tilde c}, \Psi) &\rightarrow& 
(O(1),e^{-3 r})\\
(  \hat{\tilde c}, \Psi) &\rightarrow& (e^{-r},O(1)).
\label{asymptotn0}
\end{eqnarray}
Both are acceptable solutions. While for the first, the boundary term at $r=\infty$
vanishes, for the second the boundary term does not vanish. Thus for $n=0$ or $p=0$ (\ref{actionintegral}) 
does not give any useful information. 
For the co-kernel therefore, we need to study the three special cases: i) $p=0$, $n \neq 0$, ii) $p \neq 0$ and $n=0$, 
iii) $(p,n)=(0,0)$. Fortunately, for these cases one can find explicit analytic solutions for the co-kernel equations.\\

\noindent {\bf i) $p=0$ and  $n \neq 0$}\\

In this case we can solve for $\Psi$ using the third co-kernel equation.
\begin{equation}
 \Psi(r)= -\frac{\partial_r  \hat{\tilde c}(r)}{2 L n \sinh r}
\end{equation}
Substituting this in the second co-kernel equation one gets a second order differential equation which has the general solution:
\begin{equation}
\hat{\tilde c}(r)= C_1 (\cosh r)^ {\gamma_+} + C_2 (\cosh r)^ {\gamma_-}
\end{equation}
where $\gamma_{\pm}$ are defined in (\ref{gamma}). The acceptable solution is the one with $\gamma_-$ and for this the complete one-parameter family of solution, 
after solving the inhomogeneous equation for $c$, is
\begin{eqnarray}
\hat{\tilde c}(r)&=& C_1 (\cosh r)^ {\gamma_-}\\
\Psi(r)&=& -\frac{\partial_r  \hat{\tilde c}(r)}{2 L n \sinh r}\\
c(r)&=&\frac{\hat{\tilde c}(r)}{2n}
\end{eqnarray}\\

\noindent {\bf ii) $p \neq 0$ and $n=0$}\\

In this case we can solve for $\Psi$ using the third co-kernel equation.
\begin{equation}
 \Psi(r)= \frac{\sinh r \partial_r  \hat{\tilde c}(r)}{2 p}
\end{equation}
Substituting this in the second co-kernel equation one gets a second order differential equation
\begin{equation}
 \sinh r \partial_r^2  \hat{\tilde c}(r)+ \cosh r  \partial_r\hat{\tilde c}(r)- p^2 \frac{\sech^2 r}{\sinh r} \hat{\tilde c}(r)=0
\end{equation}
The indicial roots near $r=0$ are $\pm |p|$. Asymptotically, changing the variable from $r$ to $z=e^{-r}$, the indicial roots at $z=0$ are $0$ and $1$. 
Moreover expanding the two solutions 
near $z=0$ shows that there are no logarithmic terms in $z$.  The smooth solution, i.e. the solution that behaves as $r^{|p|}$ at $r=0$, 
can be expressed in terms of hypergeometric function: 
\begin{equation}
 \hat{\tilde c}(r)= (\tanh r)^{|p|}
 {}_2F_1\l(\frac{1}{2}(|p|+\hat{\gamma}_-),\frac{1}{2}
(|p|+\hat{\gamma}_+);1+|p|;\tanh^2r\r)
\end{equation}
where $\hat{\gamma}_{\pm} = \frac{1}{2}(1\pm \sqrt{1+4 p^2})$. This solution, when analytically continued to the asymptotic region, behaves as 
\begin{equation}
 \frac{\sqrt{\pi} \Gamma(1+|p|)}{\Gamma(\frac{1}{2}(1+|p|+\hat{\gamma}_-)\Gamma(\frac{1}{2}(1+|p|+\hat{\gamma}_+)} +  \frac{4\sqrt{\pi}  \Gamma(1+|p|)}
 {\Gamma(\frac{1}{2}(|p|+\hat{\gamma}_-)\Gamma(\frac{1}{2}(|p|+\hat{\gamma}_+)} e^{-r} +...
\end{equation}
so that it is a linear combination of the two asymptotic solutions that are both acceptable. Particularly note that the coefficient of solution behaving as 
$ e^{-r}$
is non-vanishing. This is consistent with the argument given in (\ref{asymptotn0}) that shows that the boundary term in (\ref{actionintegral}) at $r=\infty$ 
does not vanish for 
the solution of $(\hat{\tilde c},\Psi)$ that behaves as $(e^{-r},O(1))$.

Substituting the above solution for $(\hat{\tilde c}(r),\Psi(r))$ in the first co-kernel equation, we can easily find the inhomogeneous solution for $c$. 
The complete one-parameter solution for $n=0$ and $p \neq 0$ is:
\begin{eqnarray}
\hat{\tilde c}(r)&=& C_2 (\tanh r)^{|p|}
 {}_2F_1\l(\frac{1}{2}(|p|+\hat{\gamma}_-),\frac{1}{2}
(|p|+\hat{\gamma}_+);1+|p|;\tanh^2r\r)\\
\Psi(r)&=& -\frac{\sinh r \partial_r  \hat{\tilde c}(r)}{2 p}\\
c(r)&=&\frac{L \hat{\tilde c}(r)}{2 p}
\end{eqnarray}\\

\noindent {\bf iii) $(p,n)=(0,0)$}\\

In this case the solution for the last two co-kernel equations have constant solutions  $\hat{\tilde c}(r)= C_1$ and $\Psi=C_2$. Plugging this in the first co-kernel 
equation one finds the solution for $c$ as
\begin{equation}
 c(r) = C_3 + C_2 L \log( \tanh\frac{r}{2}) +C_4( \log( \tanh\frac{r}{2})+ \cosh r)  
\end{equation}
Smoothness at $r=0$ and the asymptotic boundary condition on $c$ implies that  $C_4= C_2=0$. Thus we have a two paraometer family of solutions :
\be\label{const.mode}
\hat{\tilde c}(r)= C_1,\,\,\Psi(r)= 0,\,\, c(r)=C_2\,.
\ee
These are, however, just the constant modes of $c$ and $\tilde{c}$ that, we had argued earlier, should be removed from the 
path-integral. 

\subsection{Summary  and  result for the one loop determinant}
Let us summarize here the results. Dim(ker($D_{10}$))=0 for all $n$ and $p$.  Dim(coker($D_{10}$))
=0 for $n\neq 0$ and $p\neq 0$,  Dim(coker($D_{10}$))
=1 for $p=0$ and $n \neq 0$ as well as for $n=0$ and $p \neq 0$ and Dim(coker($D_{10}$))=2 for $n=p=0$.  Therefore the index of $D_{10}$ 
operator is
\begin{enumerate}\label{resultforIndex}
\item \text{ind}$D_{10} =0$ for $n\neq 0$ and $p\neq 0$\,, 
\item \text{ind}$D_{10} =-1$ for $n\neq 0$ and $p=0$\,,  
\item \text{ind}$D_{10}=-1$ for $n=0$ and $p\neq 0$\,,
\item \text{ind}$D_{10}=-2$ for $n=0$ and $p=0$\,.
\end{enumerate}
Now we note that for the case $(n=0,\,p=0)$ we have two constant ghost modes (\ref{const.mode}). As we argued earlier these 
are zero modes and we take the path integral measure such that there are no integrations over constant mode for ghost fields. We therefore, neglect these zero modes and their contribution to index\,\footnote{One could absorb these zero modes by introducing two bosonic ghosts of ghost which would give a contribution of $+2$ to the index. It would be interesting to check this explicitly following the  analysis in \cite{Pestun:2016zxk}.}. \\
Combining all the results above we get the final answer for the 1-loop super-determinant around the saddle point:
\begin{equation}
 Z_{1-\text{loop}}(\alpha)= \prod_{\rho} \sqrt{\prod_{n \neq 0}(n -i \rho\cdot\alpha) 
 \prod_{p \neq 0} (\frac{p}{L} -i \rho\cdot\alpha) }. 
\end{equation}
Substituting the above in (\ref{partition.fn.}), we obtain the matrix model
\ben\label{finalans}
Z &=&\int d\alpha ~\exp(-{\pi}iL\kappa\text{Tr}\alpha^2)
\prod_{\rho>0}(\rho.\alpha)^2\prod_{\rho} \sqrt{\prod_{n \neq 0}(n -i \rho\cdot\alpha) 
\prod_{p \neq 0} (\frac{p}{L} -i \rho\cdot\alpha) }\nn\\&=&\mathcal{N}
\int d\alpha ~\exp(-{\pi}iL\kappa\text{Tr}\alpha^2)\prod_{\rho>0}\sinh(\pi\rho\cdot\alpha)\sinh(\pi L\rho\cdot\alpha). 
\een
Here $\mathcal N$ is some constant which depends only on $L$. In order to compute 
the $L$ dependence in $\mathcal N$, we need to include also the contribution to the determinant 
from the Cartan part\footnote{The Cartan part of the determinant does not contain any $\alpha$ dependence 
and hence are not useful for most of the analysis.} 
$\prod_{i=1}^{\text{rank}}\sqrt{\prod_{n \neq 0}n\prod_{p \neq 0} (\frac{p}{L}) }$. Therefore, we 
need to regularize the infinite product of $L$ coming from both the Cartan and non-Cartan part of the determinant. 
Using the zeta function regularization one finds that $\mathcal N\propto L^{\frac{r}{2}}$ where $r$ is the 
rank of the Lie algebra. Thus we find that the $L$ dependence in $\mathcal N$ exactly cancel the 
$\sqrt{L}$ dependence in (\ref{AbelianCSint.})\footnote{We note here that there might be a $L$ dependence in the Jacobian in the path integral measure coming from the change of variables to cohomological variables.}. 
In the non abelian case there is another potential 
source of $L$ in the partition function coming from the $L$ dependence in one of the $\sinh$ function. We do 
this integral explicitly for a general gauge group in the appendix C and find that total $L$ dependence in the partition function $Z$ coming from the index computation is just a pure phase. The rest of the integral equals that of the partition function of the bosonic Chern-Simons theory on S$^3$\,\footnote{Note that going from first line to second line in (\ref{finalans}), converting infinite product to the hyperbolic functions, we have ignored infinite products of $(-1)$'s under square root. Carefully treating the sign in the square root using the gauge invariant regularization as in \cite{Kallen:2011ny}, one recovers the usual shift in $\kappa$.}.  
Finally we mention the result in (\ref{finalans}) is also equal  to  the result 
of the partition function of Chern-Simons theory of a $q$ fold cover of the 
sphere  $S^3$ where $q = \frac{1}{L}$	
obtained in \cite{Nishioka:2013haa}. This can be seen by a simple rescaling of the integral in 
(\ref{finalans}). 
\section{Wilson loop }
In this section we determine the expectation value of a  supersymmetric
Wilson loop in $AdS_2\times S^1$. 
We consider the following Wilson loop operator in the representation $R$ of the gauge group
\be
W_R=\frac{1}{\text{dim}R}\text{Tr}_RP\exp[i\oint dt (A_\mu  \dot x^\mu-\sigma |\dot x| )]\,.
\ee
The susy transformation of the above Wilson loop is given by
\be
\delta W_R=\frac{1}{\text{dim}R}\text{Tr}_RP\l[i\oint dt (\delta A_\mu  \dot x^\mu-\delta\sigma |\dot x| )\r]\exp[i\oint dt (A_\mu  \dot x^\mu-\sigma |\dot x| )]\,.
\ee
Using susy transformation of the vector multiplet, we write the above expression as
\be
\delta W_R=\frac{1}{\text{dim}R}\text{Tr}_RP \oint\{\frac{i}{2}(\epsilon\gamma_\mu\tilde\lambda+\tilde\gamma_\mu\lambda)\dot x^\mu+\frac{i}{2}(\epsilon\tilde\lambda-\tilde\epsilon\lambda)|\dot x|\}\exp[i\oint dt (A_\mu  \dot x^\mu-\sigma |\dot x| )]\,.
\ee
Thus the Wilson loop preserve susy if the following relations hold
\be
(\gamma_\mu \dot x^\mu-|\dot x|)\epsilon=0,\quad (\gamma_\mu \dot x^\mu+|\dot x|)\tilde\epsilon=0\,.
\ee
Now if we choose the Wilson loop wrapping the $\tau$-direction i.e. $\dot x^\mu=h \,e^\mu_1$, $h$ is some constant, then we find the following condition on the killing spinor
\be
(\gamma_1-1)\epsilon=0,\quad (\gamma+1)\tilde\epsilon=0
\ee
Using the explicit form of the killing spinor (\ref{susy-backgrd}), we see that this condition is satisfied if $\sinh\frac{r}{2}=0$. Thus the Wilson loop preserve the killing spinor if it wraps the circle at the origin of the AdS$_2$. The expectation value of this Wilson loop is given by
\be
<W_R>=\frac{1}{\tilde Z\,{\text{dim}R}}\int d\alpha ~\text{Tr}_R\,exp[2\pi h\alpha]\,\exp(-{\pi}iL\kappa\text{Tr}\alpha^2)\prod_{\rho>0}\sinh(\pi\rho\cdot\alpha)\sinh(\pi L\rho\cdot\alpha)\,.
\ee 
where $\tilde Z$ is the partition function without $\mathcal N$.
\section{General conformal transformations}
In this section we will consider a family of manifolds which 
are conformally equivalent to AdS$_2\times S^1$ and show that the index and the 
partition function do not change. We begin with the following metric 
\be\label{conformalMetric1}
ds^2=f_s^2(r)(d\tau^2+L^2(dr^2+\sinh^2r\,d\theta^2))\,.
\ee
The metric is conformally equivalent to AdS$_2\times S^1$ and we choose the conformal factor $f_s(r)$ such that it does not change the asymptotic behaviour of all fields. This will ensure that the space of functions over which we will calculate the index does not change drastically. This implies that we take $f_s(r)$ such that it approaches $\sim\mathcal O(1)$ as $r\rightarrow 0$ and $\infty$.
Also the family of conformally equivalent manifolds, we will consider here are labelled by parameter $s$ such that $f_{s=0}(r)=1$ which corresponds to AdS$_2\times S^1$. An example of a such function which we will use in our computations is \footnote{It would be very interesting to explore the other possible choices of function $f_s(r)$ and prove that the index does not change.}
\be
f_s(r)=1-s+s\sech r\,.
\ee
We see that for $s=1$ the metric (\ref{conformalMetric1}) is that of branched $S^3$ and any other value of $s\leq 1$ corresponds to the metric which is non singular and asymptotically  AdS$_2\times S^1$. \\
The metric (\ref{conformalMetric1}) admits Killing spinors. Following a
similar  analysis presented in  appendix B, we obtain the following Killing spinors 
\be
\epsilon= e^{\frac{i\theta}{2}}\sqrt{f_s(r)}\begin{pmatrix}i\cosh(\frac{r}{2})\\\sinh(\frac{r}{2})\end{pmatrix}\,,\qquad \tilde\epsilon=e^{-\frac{i\theta}{2}}\sqrt{f_s(r)}\begin{pmatrix}\sinh(\frac{r}{2})\\i\cosh(\frac{r}{2})\end{pmatrix}\,.
\ee
These Killing spinors correspond to the Killing vector ($K^\mu=\tilde\epsilon\gamma^\mu\epsilon$) 
\be
K=\frac{\p}{\p\tau}+\frac{1}{L}\frac{\p}{\p\theta}\,.
\ee
However in the present case the background supergravity fields $A_\mu,\,V_\mu$ and $H$
acquire 
non trivial dependence on the function $f_s(r)$ to satisfy the Killing spinor equations. 
Their explicit forms are 
\ben
&&A_\tau=\frac{2f_s(r)+3\coth r\,\p_rf_s(r)}{2Lf_s(r)}\,,\quad V_\tau=\frac{f_s(r)+\coth r\,\p_rf_s(r)}{2Lf_s(r)}\,,\quad A_{r,\theta}=0\,\nn\\&&V_{r,\theta}=0,\quad H=\frac{i\p_r f_s(r)}{Lf_s^2(r)\sinh r}\,.
\een
Thus we can use this background to compute the partition function using the localization technique. We begin with the $QV$ action. The bosonic part of the $QV$ action in the present case is given by
\ben
QV_{{\rm loc}\{\text{bosonic}\}}=\int d^3x\sqrt{g}\text{Tr}\Big[\frac{1}{4}F_{\mu\nu}F^{\mu\nu}-\frac{1}{2f^2_s(r)\cosh^2r}D_\mu(f_s(r)\cosh r\,\sigma)D^\mu(f_s(r)\cosh r\,\sigma)\nn\\-\frac{1}{32}\l(G-\frac{4}{Lf_s(r)\cosh r}\sigma\r)^2\Big]\,\nn\\.
\een
The minima of the above action are the solutions to the following equations
\be
F_{\mu\nu}=0\,,\quad D_\mu(f_s(r)\cosh r\,\sigma)=0\,,\quad G=\frac{4\sigma}{Lf_s(r)\cosh r}\,.
\ee
As we described earlier, the solution of the above equation upto a gauge transformation is given by
\be
A_{\nu}=0,\quad \sigma=\frac{i\alpha}{f_s(r)\cosh r},\quad G=\frac{i\alpha}{L(f_s(r)\cosh r)^2}\,.
\ee
And thus we see that the Chern Simons action evaluated on the above background remains unchanged
\be
\exp\l(\frac{\kappa}{{4\pi}}S_{\text{C.S.}}\r)=\exp(-{\pi}iL\kappa\text{Tr}\alpha^2)\,.
\ee
Next we look for $D_{10}$ operator. It is not very hard to convince oneself that the relevant terms in the $\hat QV$ action needed for $D_{10}$ operator are the same as given in (\ref{D10equ}). Since the killing spinors now depend on the function $f_s(r)$, the explicit form of the kernel and cokernel equations will also depend on the function $f_s(r)$ through the killing spinors. We will not present here the details of these equations. To solve these equations we will follow exactly the same analysis presented in the subsection \ref{Analysis_of_Equ}. Since we have presented the analysis for $s=0$ in details, we will not repeat here the same analysis and therefore, just state the results. We find that for $s<1$, the spaces of kernel and cokernel of the $D_{10}$ operator remain unchanged compared to $s=0$ case and thus the index of $D_{10}$ is again given by (\ref{resultforIndex}). The situation becomes interesting for the $s=1$ case. In this case we find that the spaces of kernel and cokernel of the $D_{10}
$ operator are different than the one for $s=0$ case. In particular the kernel is one dimensional for every $(n,p)$ satisfying $(n>0,p>0)$ and $(n<0,p<0)$. On the other hand the space of cokernel is also one dimensional for every combination of $(n,p)$ satisfying $(n>0,p>0)$, $(n<0,p<0)$, $(n=0,p\neq 0)$ and $(n\neq 0,p=0)$. The cokernel is 2 dimensional for $(n=0,p=0)$ which are pure constant modes for ghost fields and we do not integrate over these modes. Thus we see that although the spaces of kernel and cokernel for $s=1$ case are  very different than the $s\neq 1$, the index remains same.

\section{Conclusions}

We have used the method of localization to evaluate the partition function of
Chern-Simons theory on  the non-compact space $AdS_2\times S^1$.
The radius of $AdS_2$ is $L = 1/q$ times that of the $S^1$. 
The partition function agrees precisely with that  on 
the $q$ fold cover of $S^3$ as expected from the conformal 
symmetry which relates the partition function on these spaces. 
Furthermore since the theory is topological, this partition function is equal 
to that on $S^3$ by a pure phase which depends on $L$, upto some $L$ dependent factor coming from the Jacobian in the path integral measure. 
This constitutes  a non-trivial check of the method of localization developed 
for $AdS_2\times S^1$ in this paper. 
Though this paper  focuses  on the ${\cal N}=2$ vector multiplet, 
the method can be generalised to matter multiplets and to theories with 
higher supersymmetry. 
We expect the equality  between partition functions of conformal fields theories 
on $AdS_2 \times S^1$ and $S^3$ to hold for general super conformal  field 
theories in $3$ dimensions. 

In our analysis we showed that the relation  between the partition function on 
$AdS_2\times S^1$ and $S^3$ was obtained by considering the usual 
space of square integrable wave functions on $AdS_2\times S^1$. 
The localizing Lagrangian in particular did not develop any boundary 
terms in any steps which involved a total derivative. 
The fields satisfied boundary conditions to ensure that total derivative terms  vanished
at the origin and the boundary of $AdS_2$.

Finally we mention that this method of localization developed for 
$AdS_2\times S^1$ can be generalized to higher dimensions. 
The space $AdS_2 \times S^2$ is  particularly an
interesting one. One can  extend the approach of this paper and 
address localization of  4 dimensional supersymmetric field theories in non-compact space. 
There is an added benefit of studying localization in this space. 
$AdS_2\times S^2$ is the near horizon geometry of 
supersymmetric black holes in 4 dimensions. 
Developing   localization in this space will lead to a better understanding 
 of black hole microstates from the bulk.  We hope to address some of these 
 aspects in the future.

\section*{Acknowledgements}
We thank useful conversations with Nadav Drukker, Alba Grassi, Sameer Murthy and Ashoke Sen. We would specially like to 
thank George Thompson for many valuable comments and insightful discussion during the progress of this work. 
 J.R.D would like to thank the hospitality of the Abdus Salam ICTP during  which 
 this collaboration was initiated. 
\appendix
\section{Conventions}
The covariant derivative of a fermion is given by
\be
\nabla_\mu\psi=\l(\p_\mu+\frac{i}{4}\omega_{\mu\,ab}\varepsilon^{abc}\gamma_c\r)\psi,\qquad \varepsilon^{123}=1.
\ee
Our choice of gamma matrices are
\be
\gamma^1=\begin{pmatrix}1&0\\0& -1\end{pmatrix},\quad \gamma^2=\begin{pmatrix}0&-1\\-1& 0\end{pmatrix},\quad \gamma^3=\begin{pmatrix}0&i\\-i& 0\end{pmatrix}\,.
\ee
They satisfy gamma matrices algebra
\be
\gamma^a\gamma^b=\delta^{ab}+i\varepsilon^{abc}\gamma_c\,.
\ee
\be
\gamma^{aT}=-C\gamma^aC^{-1},\quad C=\begin{pmatrix}0 & 1\\-1 & 0\end{pmatrix},\quad C^T=-C=C^{-1}
\ee
In Lorentzian space $\psi$ and $\bar\psi$ are complex conjugate to each other but in Euclidean space fermions $\psi$ and $\bar\psi$ are independent two component complex spinor. The product of two fermions $\epsilon$ and $\psi$ is defined through charge conjugation matrix
\be
\epsilon\psi=\epsilon^TC\psi\,.
\ee

\section{Solving Killing spinor equations}\label{Killingspinor}
The Killing spinor equations are given by
\ben\label{}
\l(\nabla_\mu-iA_\mu\r)\epsilon=-\frac{1}{2}H\gamma_\mu\epsilon-i V_\mu\epsilon-\frac{1}{2}\epsilon_{\mu\nu\rho}V^\nu\gamma^\rho\epsilon\nn\\
\l(\nabla_\mu+iA_\mu\r)\tilde\epsilon=-\frac{1}{2}H\gamma_\mu\tilde\epsilon+i V_\mu\tilde\epsilon+\frac{1}{2}\epsilon_{\mu\nu\rho}V^\nu\gamma^\rho\tilde\epsilon
\een
Here $\varepsilon^{\mu\nu\rho}=\frac{1}{\sqrt{g}}\epsilon^{\mu\nu\rho},\quad \epsilon^{\tau\eta\theta}=1$.\\
In order to solve the above equations we make the following ansatz
\be
\epsilon(\tau,r,\theta)= e^{\frac{i\theta}{2}}\begin{pmatrix}\epsilon_1(r)\\\epsilon_2(r)\end{pmatrix},\quad \tilde\epsilon(\tau,r,\theta)= e^{-\frac{i\theta}{2}}\begin{pmatrix}\tilde\epsilon_1(r)\\\tilde\epsilon_2(r)\end{pmatrix},\quad V_r=V_\theta=0\,.
\ee 
In particular the ansatz for Killing spinor does not depend on the $\tau$-coordinate. Solving the 
$\tau$ component equations, one finds 
\be
A_\tau=V_\tau, \quad H=0\,.
\ee
The $\theta$-component equation is given by
\be
\begin{pmatrix}1-\cosh r-2A_\theta & -i\sinh r\,V_\tau L\\-i\sinh r\,V_\tau L&1-2A_\theta+\cosh r\end{pmatrix}\begin{pmatrix}\epsilon_1(r)\\\epsilon_2(r)\end{pmatrix}=0
\ee
Similarly for $\tilde\xi$. Requiring the existence of a non trivial solution for $\xi$ determines $A_\theta$ in terms of $V_\tau$ as
\be\label{Atheta-equ}
A_\theta=\frac{1}{2}\l(1\pm\sqrt{1+(1-L^2V_\tau^2)\sinh^2r}\r)
\ee
Now we look at the $r$-component equations,
\ben\label{r-derivativeq}
&&\p_r\epsilon_1(r)-iA_r\epsilon_1(r)-\frac{i}{2}LV_\tau\epsilon_2(r)=0\nn\\
&&\p_r\epsilon_2(r)-iA_r\epsilon_2(r)+\frac{i}{2}LV_\tau\epsilon_1(r)=0\nn\\
&&\p_r\tilde\epsilon_1(r)+iA_r\tilde\epsilon_1(r)+\frac{i}{2}LV_\tau\tilde\epsilon_2(r)=0\\
&&\p_r\tilde\epsilon_2(r)+iA_r\tilde\epsilon_2(r)-\frac{i}{2}LV_\tau\tilde\epsilon_1(r)=0\nn
\een 
One finds that if we define $R=\frac{\epsilon_2(r)}{\epsilon_1(r)}$ and $\tilde R=\frac{\tilde\epsilon_1(r)}{\tilde\epsilon_2(r)}$, then from above set of equations
\be\label{R-equ}
\p_rR=-\frac{iL}{2}V_\tau(1+R^2),\quad \p_r\tilde R=-\frac{iL}{2}V_\tau(1+\tilde R^2)\,.
\ee
Now let us look at the form of the Killing vector.
\be
K^\mu=\tilde\epsilon\gamma^\mu\epsilon=(a,0,b)
\ee
\ben
\tilde\epsilon_1(r)\epsilon_2(r)+\tilde\epsilon_2(r)\epsilon_1(r)=-a,\quad \tilde\epsilon_1(r)\epsilon_1(r)=\tilde\epsilon_2(r)\epsilon_2(r)=\frac{ibL}{2}\sinh r
\een
Using the above equations it is very simple to determine $R$ which is given by
\be
R=\frac{\epsilon_2(r)}{\epsilon_1(r)}=i\frac{a\pm\sqrt{a^2+b^2L^2\sinh^2r}}{bL\sinh r}
\ee
Substituting above in (\ref{R-equ}), we determine $V_\tau$ as
\be
V_\tau=\mp\frac{b\cosh r}{\sqrt{a^2+b^2L^2\sinh^2r}}
\ee
Substituting the expression of $V_\tau$ in (\ref{Atheta-equ}), we obtain
\be
A_\theta=\frac{1}{2}\l[1\pm\frac{a\cosh r}{\sqrt{a^2+b^2L^2\sinh^2r}}\r]
\ee
Thus susy preserving backgrounds are labelled by two real parametrs. Our choice of susy transformation parameters (\ref{susy-backgrd}) satisfy above equations with the choice $a=1,\, b=\frac{1}{L}$.  
\section{$L$ dependence in the partition function}
To demonstrate that the $L$ dependence in the partition function is 
just a pure phase we evaluate the integral
\be
\tilde Z=\int \prod_{i=1}^rd\alpha_i\, e^{-\pi iL\kappa\text{Tr}\alpha^2}\prod_{\rho>0}\sinh(\pi\rho\cdot\alpha)\sinh(\pi L\rho\cdot\alpha)
\ee
This is done 
 following the steps given in \cite{Aganagic:2002wv}.  First we will use Weyl denominator formula
\be
\sum_{w\in\mathcal W}\epsilon(w)\,e^{w(\delta)\cdot \alpha}=\prod_{\rho>0}2\sinh(\frac{\rho\cdot\alpha}{2})
\ee 
Here $\delta$ is the sum over positive roots
\be
\delta=\frac{1}{2}\sum_{\rho>0}\rho
\ee
We get
\ben
\tilde Z&=&\frac{1}{L^{r/2}}\int \prod_{i=1}^rd\mu_i\, e^{-\frac{1}{2g_s}\text{Tr}\mu^2}\prod_{\rho>0}\sinh(\frac{\pi\rho\cdot\mu}{\sqrt{L}})\sinh(\pi \sqrt{L}\rho\cdot\mu)\,,\nn\\&&=\frac{1}{2^{2\Delta_+}L^{r/2}}\int \prod_{i=1}^rd\mu_i\, e^{-\frac{1}{2g_s}\text{Tr}\mu^2}\sum_{w,w'\in\mathcal W}\epsilon(w'')\,e^{w(\delta)\cdot \frac{\pi\mu}{\sqrt{L}}}e^{w'(\delta)\cdot \pi\mu\sqrt{L}}\,.
\een
Here $\sqrt{L}\alpha=\mu,\quad\frac{1}{2g_s}=\pi i\kappa,\quad w''=w\cdot w'$ and $\Delta_+=$ total number of positive roots. Now we can explicitly do the above Gaussian integral. 
\ben
\tilde Z&=&\frac{(\text{det}C)^{1/2}}{2^{2\Delta_+}}\l(\frac{2\pi g_s}{L}\r)^{r/2}\sum_{w,w'\in \mathcal W}\epsilon(w'') e^{\frac{g_s\pi^2}{2}(\frac{w(\delta)}{\sqrt{L}}+w'(\delta)\sqrt{L})\cdot(\frac{w(\delta)}{\sqrt{L}}+w'(\delta)\sqrt{L})}\,,\nn\\&&=\frac{(\text{det}C)^{1/2}}{2^{2\Delta_+}}\l(\frac{2\pi g_s}{L}\r)^{r/2}|\mathcal W|\sum_{w''\in \mathcal W}\epsilon(w'')e^{\frac{g_s\pi^2}{2}(\delta\cdot\delta)(\frac{1}{L}+L)}\,e^{g_s\pi^2(\delta\cdot w''(\delta))}\,,\nn\\&&=\frac{(\text{det}C)^{1/2}}{2^{2\Delta_+}}\l(\frac{2\pi g_s}{L}\r)^{r/2}|\mathcal W|e^{\frac{g_s\pi^2}{2}(\delta\cdot\delta)(\frac{1}{L}+L)}\prod_{\rho>0}2\sinh(\frac{g_s\pi^2\rho\cdot\delta}{2})\,.
\een
Here $C$ and $|\mathcal W|$ are the inverse of Cartan matrix and the order of the Weyl group, respectively. Now using Freudenthal de Vries formula
\be
(\delta\cdot\delta)=\frac{d_G\,y}{12}\,,
\ee 
where $d_G$ is the dimension of the group and $y$ is the dual Coxeter number, we get
\be
\tilde Z=\frac{(\text{det}C)^{1/2}}{2^{\Delta_+}}\l(\frac{2\pi g_s}{L}\r)^{r/2}|\mathcal W|e^{-\frac{i\pi\Delta_+}{2}}e^{-i\frac{\pi}{48\kappa}d_Gy(\frac{1}{L}+L)}\prod_{\rho>0}\sin(\frac{\pi}{4\kappa}\rho\cdot\delta)
\ee
Thus we see that the entire $L$ dependence is a phase plus overall $\frac{1}{L^{r/2}}$.
This together with 
 the $L^{r/2}$ contribution in $\mathcal N$ in (\ref{finalans})  leaves behind a partition function 
 which  depends on the metric through a pure phase. 



\begin{thebibliography}{10}

\bibitem{Witten:1992xu}
E.~Witten, {\it {Two-dimensional gauge theories revisited}},  {\em J. Geom.
  Phys.} {\bf 9} (1992) 303--368,
  [\href{http://arxiv.org/abs/hep-th/9204083}{{\tt hep-th/9204083}}].

\bibitem{Nekrasov:2003af}
N.~A. Nekrasov, {\it {Seiberg-Witten prepotential from instanton counting}},
  in {\em {International Congress of Mathematicians (ICM 2002) Beijing, China,
  August 20-28, 2002}}, 2003.
\newblock \href{http://arxiv.org/abs/hep-th/0306211}{{\tt hep-th/0306211}}.

\bibitem{Nekrasov:2003rj}
N.~Nekrasov and A.~Okounkov, {\it {Seiberg-Witten theory and random
  partitions}},  {\em Prog. Math.} {\bf 244} (2006) 525--596,
  [\href{http://arxiv.org/abs/hep-th/0306238}{{\tt hep-th/0306238}}].

\bibitem{Pestun:2007rz}
V.~Pestun, {\it {Localization of gauge theory on a four-sphere and
  supersymmetric Wilson loops}},  {\em Commun. Math. Phys.} {\bf 313} (2012)
  71--129, [\href{http://arxiv.org/abs/0712.2824}{{\tt arXiv:0712.2824}}].

\bibitem{Pestun:2016zxk}
V.~Pestun et~al., {\it {Localization techniques in quantum field theories}},
  \href{http://arxiv.org/abs/1608.0295}{{\tt arXiv:1608.0295}}.

\bibitem{Festuccia:2011ws}
G.~Festuccia and N.~Seiberg, {\it {Rigid Supersymmetric Theories in Curved
  Superspace}},  {\em JHEP} {\bf 06} (2011) 114,
  [\href{http://arxiv.org/abs/1105.0689}{{\tt arXiv:1105.0689}}].

\bibitem{Dumitrescu:2012ha}
T.~T. Dumitrescu, G.~Festuccia, and N.~Seiberg, {\it {Exploring Curved
  Superspace}},  {\em JHEP} {\bf 08} (2012) 141,
  [\href{http://arxiv.org/abs/1205.1115}{{\tt arXiv:1205.1115}}].

\bibitem{Klare:2013dka}
C.~Klare and A.~Zaffaroni, {\it {Extended Supersymmetry on Curved Spaces}},
  {\em JHEP} {\bf 10} (2013) 218, [\href{http://arxiv.org/abs/1308.1102}{{\tt
  arXiv:1308.1102}}].

\bibitem{Alday:2013lba}
L.~F. Alday, D.~Martelli, P.~Richmond, and J.~Sparks, {\it {Localization on
  Three-Manifolds}},  {\em JHEP} {\bf 10} (2013) 095,
  [\href{http://arxiv.org/abs/1307.6848}{{\tt arXiv:1307.6848}}].

\bibitem{Closset:2013vra}
C.~Closset, T.~T. Dumitrescu, G.~Festuccia, and Z.~Komargodski, {\it {The
  Geometry of Supersymmetric Partition Functions}},  {\em JHEP} {\bf 01} (2014)
  124, [\href{http://arxiv.org/abs/1309.5876}{{\tt arXiv:1309.5876}}].

\bibitem{Aharony:2015hix}
O.~Aharony, M.~Berkooz, A.~Karasik, and T.~Vaknin, {\it {Supersymmetric field
  theories on AdS$_{p} \times$ S$^{q}$}},  {\em JHEP} {\bf 04} (2016) 066,
  [\href{http://arxiv.org/abs/1512.0469}{{\tt arXiv:1512.0469}}].

\bibitem{Dabholkar:2010uh}
A.~Dabholkar, J.~Gomes, and S.~Murthy, {\it {Quantum black holes, localization
  and the topological string}},  {\em JHEP} {\bf 06} (2011) 019,
  [\href{http://arxiv.org/abs/1012.0265}{{\tt arXiv:1012.0265}}].

\bibitem{Dabholkar:2011ec}
A.~Dabholkar, J.~Gomes, and S.~Murthy, {\it {Localization \& Exact
  Holography}},  {\em JHEP} {\bf 04} (2013) 062,
  [\href{http://arxiv.org/abs/1111.1161}{{\tt arXiv:1111.1161}}].

\bibitem{Gupta:2012cy}
R.~K. Gupta and S.~Murthy, {\it {All solutions of the localization equations
  for N=2 quantum black hole entropy}},  {\em JHEP} {\bf 02} (2013) 141,
  [\href{http://arxiv.org/abs/1208.6221}{{\tt arXiv:1208.6221}}].

\bibitem{Dabholkar:2014ema}
A.~Dabholkar, J.~Gomes, and S.~Murthy, {\it {Nonperturbative black hole entropy
  and Kloosterman sums}},  {\em JHEP} {\bf 03} (2015) 074,
  [\href{http://arxiv.org/abs/1404.0033}{{\tt arXiv:1404.0033}}].

\bibitem{Gupta:2015gga}
R.~K. Gupta, Y.~Ito, and I.~Jeon, {\it {Supersymmetric Localization for BPS
  Black Hole Entropy: 1-loop Partition Function from Vector Multiplets}},  {\em
  JHEP} {\bf 11} (2015) 197, [\href{http://arxiv.org/abs/1504.0170}{{\tt
  arXiv:1504.0170}}].

\bibitem{Murthy:2015yfa}
S.~Murthy and V.~Reys, {\it {Functional determinants, index theorems, and exact
  quantum black hole entropy}},  {\em JHEP} {\bf 12} (2015) 028,
  [\href{http://arxiv.org/abs/1504.0140}{{\tt arXiv:1504.0140}}].

\bibitem{Dabholkar:2014wpa}
A.~Dabholkar, N.~Drukker, and J.~Gomes, {\it {Localization in supergravity and
  quantum $AdS_4/CFT_3$ holography}},  {\em JHEP} {\bf 10} (2014) 90,
  [\href{http://arxiv.org/abs/1406.0505}{{\tt arXiv:1406.0505}}].

\bibitem{Casini:2011kv}
H.~Casini, M.~Huerta, and R.~C. Myers, {\it {Towards a derivation of
  holographic entanglement entropy}},  {\em JHEP} {\bf 05} (2011) 036,
  [\href{http://arxiv.org/abs/1102.0440}{{\tt arXiv:1102.0440}}].

\bibitem{Huang:2014gca}
X.~Huang, S.-J. Rey, and Y.~Zhou, {\it {Three-dimensional SCFT on conic space
  as hologram of charged topological black hole}},  {\em JHEP} {\bf 03} (2014)
  127, [\href{http://arxiv.org/abs/1401.5421}{{\tt arXiv:1401.5421}}].

\bibitem{Huang:2014pda}
X.~Huang and Y.~Zhou, {\it {$ \mathcal{N}=4 $ Super-Yang-Mills on conic space
  as hologram of STU topological black hole}},  {\em JHEP} {\bf 02} (2015) 068,
  [\href{http://arxiv.org/abs/1408.3393}{{\tt arXiv:1408.3393}}].

\bibitem{Klebanov:2011uf}
I.~R. Klebanov, S.~S. Pufu, S.~Sachdev, and B.~R. Safdi, {\it {Renyi Entropies
  for Free Field Theories}},  {\em JHEP} {\bf 04} (2012) 074,
  [\href{http://arxiv.org/abs/1111.6290}{{\tt arXiv:1111.6290}}].

\bibitem{Nishioka:2013haa}
T.~Nishioka and I.~Yaakov, {\it {Supersymmetric Renyi Entropy}},  {\em JHEP}
  {\bf 10} (2013) 155, [\href{http://arxiv.org/abs/1306.2958}{{\tt
  arXiv:1306.2958}}].

\bibitem{Cabo-Bizet:2014nia}
A.~Cabo-Bizet, E.~Gava, V.~I. Giraldo-Rivera, M.~N. Muteeb, and K.~S. Narain,
  {\it {Partition Function of $N=2$ Gauge Theories on a Squashed $S^4$ with
  $SU(2)\times U(1)$ Isometry}},  {\em Nucl. Phys.} {\bf B899} (2015) 149--164,
  [\href{http://arxiv.org/abs/1412.6826}{{\tt arXiv:1412.6826}}].

\bibitem{Banerjee:2010qc}
S.~Banerjee, R.~K. Gupta, and A.~Sen, {\it {Logarithmic Corrections to Extremal
  Black Hole Entropy from Quantum Entropy Function}},  {\em JHEP} {\bf 03}
  (2011) 147, [\href{http://arxiv.org/abs/1005.3044}{{\tt arXiv:1005.3044}}].

\bibitem{Camporesi:1994ga}
R.~Camporesi and A.~Higuchi, {\it {Spectral functions and zeta functions in
  hyperbolic spaces}},  {\em J. Math. Phys.} {\bf 35} (1994) 4217--4246.
 \bibitem{Kallen:2011ny} 
  J.~Kallen,
  {\it{Cohomological localization of Chern-Simons theory}},
  {\em JHEP} {\bf 1108}, 008 (2011)
   [\href{http://arxiv.org/abs/1104.5353}{{\tt arXiv:1104.5353}}].
\bibitem{Aganagic:2002wv} 
  M.~Aganagic, A.~Klemm, M.~Marino and C.~Vafa,
  {\it{Matrix model as a mirror of Chern-Simons theory}},
  {\em JHEP} {\bf 0402}, 010 (2004), 
   [\href{http://arxiv.org/abs/hep-th/0211098}{{\tt hep-th/0211098}}].

\end{thebibliography}

\providecommand{\href}[2]{#2}\begingroup\raggedright\endgroup

\end{document}